\RequirePackage{fix-cm}
\documentclass[smallextended]{svjour3wide}       
\smartqed  
\usepackage{graphicx}
\usepackage{amsmath}
\usepackage{amssymb}
\usepackage{mathrsfs}
\usepackage{color}
\usepackage{bm}
\usepackage{cite} 
\journalname{Journal of Statistical Physics}
\begin{document}

\title{Violation of the second fluctuation-dissipation relation and entropy production in nonequilibrium medium
}


\author{Tomohiro Tanogami
}


\institute{T. Tanogami \at
              Department of Physics, Kyoto University, Kyoto 606-8502, Japan \\
              \email{tanogami.tomohiro.84c@st.kyoto-u.ac.jp}           
}

\date{Received: date / Accepted: date}

\maketitle

\begin{abstract}
We investigate a class of nonequilibrium media described by Langevin dynamics that satisfies the local detailed balance.
For the effective dynamics of a probe immersed in the medium, we derive an inequality that bounds the violation of the second fluctuation-dissipation relation (FDR).
We also discuss the validity of the effective dynamics.
In particular, we show that the effective dynamics obtained from nonequilibrium linear response theory is consistent with that obtained from a singular perturbation method.
As an example of these results, we propose a simple model for a nonequilibrium medium in which the particles are subjected to potentials that switch stochastically.
For this model, we show that the second FDR is recovered in the fast switching limit, although the particles are out of equilibrium.
\end{abstract}


\section{Introduction}
The properties of a system can be investigated by observing the response of the system against external stimuli.
The first fluctuation-dissipation relation (FDR) states that, for equilibrium systems, the same information as such a response is carried by an equilibrium correlation function \cite{kubo1966fluctuation,kubo2012statistical}.
By contrast, in nonequilibrium systems, the first FDR is violated.
Even for this case, there are phenomenological relations that connect the violation of the first FDR to energy dissipation \cite{harada2005phenomenological,harada2005equality,harada2006energy,harada2007fluctuations,cugliandolo1997fluctuation}.
In particular, the Harada-Sasa equality \cite{harada2005phenomenological,harada2005equality,harada2006energy,harada2007fluctuations} enables us to measure energy dissipation from experimentally accessible quantities and has been applied to various systems from molecular motors \cite{toyabe2015single,ariga2018nonequilibrium} to turbulence \cite{matsumoto2014response,matsumoto2021correlation}.

These phenomenological relations that extend the first FDR to nonequilibrium systems are based on the second FDR, which expresses the balance between the friction and noise intensity in the sense that they are compatible with equilibrium statistics.
The second FDR requires the assumption that the nonequilibrium condition imposed on the system does not directly affect the environments, i.e., the environments are quickly equilibrated \cite{harada2006energy,maes2014second}.
Indeed, it can be derived by imposing the local detailed balance (LDB) condition \cite{hayashi2006linear,harada2006energy,maes2013fluctuation,maes2014second,maes2020response,maes2021local}.
Therefore, the second FDR can be violated if the environment itself is out of equilibrium.
Such a nonequilibrium environment can be found in various systems, particularly biological systems \cite{mizuno2007nonequilibrium,nishizawa2017feedback,jee2018catalytic}.
Nonequilibrium fluctuations generated by these environments can induce a variety of rich phenomena that cannot be found in equilibrium systems.
For example, the speeds of cargos transported by kinesin in cells are much faster than \textit{in vitro} although the cell interior is crowded and viscous \cite{ross2016dark}.
In this regard, Ariga \textit{et al.}\ have recently shown that kinesin is accelerated by nonthermal fluctuations \cite{ariga2021noise}.
It is thus desirable to characterize and classify nonequilibrium environments to deepen our understanding of the phenomena induced by nonequilibrium fluctuations.

As a first step toward this end, we investigate a simple class of nonequilibrium media and seek universal relations on the violation of the second FDR.
Specifically, we consider a system consisting of three levels of description: probe, driven particles (nonequilibrium medium), and equilibrium thermal bath.
We focus on a class of nonequilibrium media described by Langevin dynamics that satisfies the LDB.
Such a formulation has been used in several works to investigate the effective dynamics of a probe immersed in nonequilibrium media \cite{hayashi2006law,maes2014second,maes2015friction,haga2015nonequilibrium,steffenoni2016interacting,maes2017induced}.
For this setup, we derive the effective dynamics of the probe by using nonequilibrium linear response theory \cite{baiesi2009fluctuations,baiesi2009nonequilibrium,baiesi2013update,maes2014second,maes2015friction,steffenoni2016interacting,maes2017induced,maes2020fluctuating,maes2020response} and investigate the violation of the second FDR.

In this paper, we derive an inequality that bounds the violation of the second FDR.
This inequality states that the violation of the second FDR is bounded by the fluctuation of the ``response'' of the total stochastic entropy production in the nonequilibrium medium against a perturbation of the probe position.
We also discuss the validity of the effective dynamics.
In particular, we show that the effective dynamics obtained from a singular perturbation method corresponds to that obtained from nonequilibrium linear response theory in the Markovian limit.
As a simple example of these results, we introduce a \textit{potential switching medium}, the particles of which are described by the so-called \textit{potential switching model}, i.e., overdamped Langevin dynamics with a stochastically switching potential \cite{toyabe2007experimental,dieterich2015single,wang2016entropy}.
For this simple linear system, all relevant quantities can be calculated explicitly.
We show that the standard second FDR is recovered in the fast switching limit, although the driven particles are out of equilibrium because of the so-called \textit{hidden entropy} \cite{celani2012anomalous,kawaguchi2013fluctuation,wang2016entropy}.
Correspondingly, we show that the upper bound of the inequality for the violation of the second FDR goes to zero in this limit.

This paper is organized as follows.
In Sect.\ \ref{Setup}, we explain the setup.
In Sect.\ \ref{Effective dynamics and the bound on the violation of the second FDR}, we present the effective dynamics of the probe, in which the second FDR is violated in general.
Then, we explain the inequality that bounds the violation of the second FDR, which is our first main result.
In Sect.\ \ref{Derivation}, we review the derivation of the effective dynamics based on nonequilibrium linear response theory.
Then, we derive the inequality for the violation of the second FDR.
The validity of the effective dynamics is discussed in Sect.\ \ref{Validity of the effective dynamics}.
We show that the effective dynamics is consistent with the result obtained by using a singular perturbation method.
In Sect.\ \ref{Example}, we introduce the potential switching medium as a simple example.
Concluding remarks are provided in Sect.\ \ref{Concluding remarks}.

\section{Setup\label{Setup}}
In this section, we explain the setup, which consists of three levels of description: probe, driven particles (nonequilibrium medium), and equilibrium thermal bath.
While we consider one-dimensional systems in the following, the results can be generalizable to higher dimensions.
Let $X_t$ be the position of a probe with mass $M$ at time $t$.
The probe is in contact with both an equilibrium thermal bath at temperature $T$ and a nonequilibrium medium that consists of $N$ particles, the positions of which are denoted by $x^j_t$ ($j=1,2,\cdots,N$).
We denote the collection of $x^j$ as ${\bm x}:=\{x^1,x^2,\cdots,x^N\}$.
The time evolution of $X_t$ is given by the following underdamped Langevin equation:
\begin{align}
M\ddot{X}_t=\Phi({\bm x}_t,X_t)-\Gamma\dot{X}_t+\sqrt{2B}\Xi_t.
\label{Model_probe}
\end{align}
Here, $\Phi({\bm x}_t,X_t)$ represents the interaction force between the probe and the particles described by the coupling potential $V({\bm x},X)$:
\begin{align}
\Phi({\bm x}_t,X_t):=-\lambda\dfrac{\partial}{\partial X_t}V({\bm x}_t,X_t),
\end{align}
where $\lambda$ denotes the dimensionless coupling constant, which can be scaled with $N$.
The second and third terms on the right-hand side of (\ref{Model_probe}) represent the coupling with the equilibrium thermal bath, where $\Gamma$ denotes the friction coefficient and $\Xi_t$ is the zero-mean white Gaussian noise that satisfies
\begin{equation}
\langle\Xi_t\Xi_s\rangle=\delta(t-s).
\end{equation}
The noise intensity $B$ is related to the friction coefficient $\gamma$ and temperature $T$ through the second FDR: $B=\Gamma k_{\mathrm{B}}T$.

The dynamics of the particles is described by the following overdamped Langevin equation:
\begin{align}
\gamma\dot{x}^j_t=F^j({\bm x}_t)-\lambda\dfrac{\partial}{\partial x^j_t}V({\bm x}_t,X_t)+\sqrt{2B_m}\xi^j_t.
\label{Model_particles}
\end{align}
Here, $F^j({\bm x})$ denotes the force acting on the $j$-th particle, generally consisting of nonconservative forces and interactions between particles.
The second term on the right-hand side of (\ref{Model_particles}) represents the interaction with the probe.
The last term represents the thermal noise: $\xi^j_t$ is the zero-mean white Gaussian noise that satisfies
\begin{align}
\langle\xi^i_t\xi^j_s\rangle=\delta_{ij}\delta(t-s),
\end{align}
and the noise intensity $B_m$ is related to $\gamma$ and $T$ through the second FDR: $B_m=\gamma k_{\mathrm{B}}T$.
We remark that the following results are valid even in the case where $F^j({\bm x})$ includes additional random forces as long as the LDB is satisfied for the additional degrees of freedom (see, e.g., Sect.\ \ref{Example}).

We are interested in the regime where the motion of the probe is much slower than that of the particles so that the probe dynamics can be described by some effective model.
This assumption will be described more explicitly in the singular perturbation method described in Sect.\ \ref{Validity of the effective dynamics}.

\section{Effective dynamics and the bound on the violation of the second FDR\label{Effective dynamics and the bound on the violation of the second FDR}}
Under the setup described in Sect.\ \ref{Setup}, we can derive the effective dynamics of the probe by eliminating the degrees of freedom of the nonequilibrium medium.
The resulting effective dynamics does not generally satisfy the second FDR.
In this section, we summarize the effective dynamics of the probe and present our first main result on the violation of the second FDR.
The derivation of these results is provided in the next section.

\subsection{Effective dynamics of the probe}
The effective dynamics of the probe is described by the following generalized Langevin-type equation:
\begin{align}
M\ddot{X}_t=G(X_t)-\Gamma\dot{X}_t-\int^t_{-\infty}ds\gamma(t-s)\dot{X}_s+\sqrt{2\Gamma k_{\mathrm{B}}T}\Xi_t+\eta_t.
\label{Effective generalized Langevin}
\end{align}
This result states that the interaction force $\Phi$ is decomposed into three parts: the streaming term $G(X_t)$, the friction force with the memory kernel $\gamma(t-s)$, and the zero-mean colored noise $\eta_t$.
The streaming term $G(X_t)$ is given by
\begin{align}
G(X_t):=\langle\Phi({\bm x}_t,X_t)\rangle^{X_t},
\end{align}
where $\langle\cdot\rangle^{X_t}$ denotes the average with respect to the stationary distribution $P^{X_t}_{\mathrm{ss}}({\bm x})$ for the particle dynamics (\ref{Model_particles}) with $X_t$ held fixed.
The friction kernel is given by
\begin{align}
\gamma(t-s):=\dfrac{1}{2k_{\mathrm{B}}T}\int^s_{-\infty}du\left[\dfrac{d}{du}\langle \Phi({\bm x}_u,X_t);\Phi({\bm x}_t,X_t)\rangle^{X_t}-\langle\mathcal{L}^\dagger_u\Phi({\bm x}_u,X_t);\Phi({\bm x}_t,X_t)\rangle^{X_t}\right],\quad\text{for}\quad t\ge s.
\label{friction kernel}
\end{align}
Here, $\langle f;g\rangle^{X_t}:=\langle fg\rangle^{X_t}-\langle f\rangle^{X_t}\langle g\rangle^{X_t}$ and $\mathcal{L}^\dagger_u$ denotes the backward operator of the dynamics (\ref{Model_particles}) with $X_t$ held fixed:
\begin{align}
\mathcal{L}^\dagger_u:=\sum_j\left[\dfrac{1}{\gamma}\left(F^j({\bm x}_u)-\lambda\dfrac{\partial}{\partial x^j_u}V({\bm x}_u,X_t)\right)\dfrac{\partial}{\partial x^j_u}+\dfrac{k_{\mathrm{B}}T}{\gamma}\dfrac{\partial^2}{\partial (x^j_u)^2}\right].
\label{back ward operator}
\end{align}
The noise term is expressed as $\eta_t=\Phi({\bm x}_t,X_t)-\langle\Phi({\bm x}_t,X_t)\rangle^{X_t}$, and its correlation is related to the friction kernel $\gamma(t-s)$ in the following form: 
\begin{align}
\langle\eta_t\eta_s\rangle^{X_t}=k_{\mathrm{B}}T\left[\gamma(t-s)+\gamma_{\mathrm{ex}}(t-s)\right],
\end{align}
where $\gamma_{\mathrm{ex}}(t-s)$ denotes the excess friction kernel defined as
\begin{align}
\gamma_{\mathrm{ex}}(t-s):=\dfrac{1}{2k_{\mathrm{B}}T}\int^s_{-\infty}du\left[\dfrac{d}{du}\langle \Phi({\bm x}_u,X_t);\Phi({\bm x}_t,X_t)\rangle^{X_t}+\langle\mathcal{L}^\dagger_u\Phi({\bm x}_u,X_t);\Phi({\bm x}_t,X_t)\rangle^{X_t}\right].
\end{align} 
Because of this excess friction kernel, the second FDR is generally violated.
In the equilibrium case, however, one can immediately confirm that the standard second FDR holds:
from the time-reversal symmetry,
\begin{align}
\langle\mathcal{L}^\dagger_u\Phi({\bm x}_u,X_t);\Phi({\bm x}_t,X_t)\rangle^{X_t}_{\mathrm{eq}}&=\langle\mathcal{L}^\dagger_t\Phi({\bm x}_t,X_t);\Phi({\bm x}_u,X_t)\rangle^{X_t}_{\mathrm{eq}}\notag\\
&=\dfrac{d}{dt}\langle \Phi({\bm x}_t,X_t);\Phi({\bm x}_u,X_t)\rangle^{X_t}_{\mathrm{eq}}\notag\\
&=-\dfrac{d}{du}\langle \Phi({\bm x}_u,X_t);\Phi({\bm x}_t,X_t)\rangle^{X_t}_{\mathrm{eq}},\quad\text{for}\quad t\ge u,
\end{align}
and thus $\gamma_{\mathrm{ex}}(t-s)=0$.
Note that the noise $\eta_t$ need not be Gaussian nor white.
The Gaussian noise may be obtained by taking the limit $N\rightarrow\infty$ combined with the weak coupling limit $\lambda\rightarrow0$ \cite{zwanzig1973nonlinear,zwanzig2001nonequilibrium,maes2020fluctuating}.

In the Markov approximation, the friction kernel is approximated as
\begin{align}
\gamma(t-s)=2\gamma_{\mathrm{eff}}\delta(t-s),
\end{align}
where $\gamma_{\mathrm{eff}}$ denotes the effective friction coefficient:
\begin{align}
\gamma_{\mathrm{eff}}:=\int^\infty_0dt\gamma(t).
\end{align}
Similarly, the excess friction kernel becomes
\begin{align}
\gamma_{\mathrm{ex}}(t-s)=2\gamma_{\mathrm{ex}}\delta(t-s)
\end{align}
with
\begin{align}
\gamma_{\mathrm{ex}}:=\int^\infty_0dt\gamma_{\mathrm{ex}}(t).
\end{align}
Thus, in the Markov approximation, the effective generalized Langevin-type equation (\ref{Effective generalized Langevin}) becomes
\begin{align}
M\ddot{X}_t=G(X_t)-(\Gamma+\gamma_{\mathrm{eff}})\dot{X}_t+\sqrt{2(\Gamma+\gamma_{\mathrm{eff}}+\gamma_{\mathrm{ex}})k_{\mathrm{B}}T}\Xi_t.
\label{Effective Langevin}
\end{align}

\subsection{Bound on the violation of the second FDR}
Let $\Delta s^{X_t}_{\mathrm{tot}}$ be the total stochastic entropy production up to time $s$ of the nonequilibrium medium in the nonequilibrium steady state (NESS) with $X_t$ held fixed.
The first main result of this paper is the following inequality, which connects the violation of the second FDR with the entropy production of the nonequilibrium medium:
\begin{align}
\left|\gamma(t-s)-\dfrac{1}{k_{\mathrm{B}}T}\langle\eta_t\eta_s\rangle^{X_t}\right|\le\sqrt{\langle\Phi^2\rangle^{X_t}}\sqrt{\mathrm{Var}\left[\partial_{X_t}\Delta s^{X_t}_{\mathrm{tot}}\right]}.
\label{bound on the violation of the 2nd FDR}
\end{align}
Here, $\mathrm{Var}[\cdot]$ denotes the variance with respect to the stationary distribution $P^{X_t}_{\mathrm{ss}}$.

If we interpret $\partial_{X_t}\Delta s^{X_t}_{\mathrm{tot}}$ as the ``response'' of the total stochastic entropy production of the nonequilibrium medium to a perturbation of the probe position, (\ref{bound on the violation of the 2nd FDR}) states that the violation of the second FDR is bounded by the fluctuation of the ``response.''
Hence, if the total stochastic entropy production is ``robust'' against the perturbation, i.e., $\mathrm{Var}[\partial_{X_t}\Delta s^{X_t}_{\mathrm{tot}}]=0$, the standard second FDR is recovered.
In particular, we can easily see that the standard second FDR holds in the equilibrium case because $\Delta s^{X_t}_{\mathrm{tot}}=0$.

\section{Derivation\label{Derivation}}
In this section, we derive the results presented in Sect.\ \ref{Effective dynamics and the bound on the violation of the second FDR}.
In particular, we review the derivation of the effective probe dynamics based on nonequilibrium linear response theory \cite{baiesi2009fluctuations,baiesi2009nonequilibrium,baiesi2013update,maes2014second,maes2015friction,steffenoni2016interacting,maes2017induced,maes2020fluctuating,maes2020response}.
In Sect.\ \ref{Validity of the effective dynamics}, we discuss the validity of this approach and show that the effective dynamics obtained from nonequilibrium linear response theory is consistent with that obtained from a singular perturbation method.

\subsection{Derivation of the effective dynamics based on nonequilibrium linear response theory}
Since we are now interested in the regime where the motion of the probe is much slower than that of the particles, we regard the probe motion as a time-dependent perturbation on the particle dynamics.
In other words, we regard the dynamics (\ref{Model_particles}) with $X_t$ held fixed as the unperturbed dynamics:
\begin{align}
\gamma\dot{x}^j_s&=F^j({\bm x}_s)-\lambda\dfrac{\partial}{\partial x^j_s}V({\bm x}_s,X_t)+\sqrt{2\gamma k_{\mathrm{B}}T}\xi^j_s,\quad\text{for}\quad s\le t.
\label{Model_particles_unperturbed}
\end{align}
Let $\mathbb{P}([{\bm x}]|X_t)$ be the probability density of a trajectory $[{\bm x}]:=\{{\bm x}_s|s\le t\}$ of this unperturbed dynamics.
Similarly, let $\mathbb{P}([{\bm x}]|[X])$ be the probability density of the original dynamics (\ref{Model_particles}) conditioned on an arbitrary probe trajectory up to time $t$, $[X]:=\{X_s|s\le t\}$.
To investigate the response of particles against the probe motion, we first compare $\mathbb{P}([{\bm x}]|[X])$ and $\mathbb{P}([{\bm x}]|X_t)$.
From the Girsanov formula \cite{girsanov1960transforming}, $\mathbb{P}([{\bm x}]|[X])$ is related to $\mathbb{P}([{\bm x}]|X_t)$ in the following form:
\begin{align}
\mathbb{P}([{\bm x}]|[X])=\exp(-\mathcal{A}([{\bm x}]|[X]))\mathbb{P}([{\bm x}]|X_t).
\label{Girsanov}
\end{align}
Here, the excess action $\mathcal{A}([{\bm x}]|[X])$ is given by
\begin{align}
-\mathcal{A}([{\bm x}]|[X])&:=\sum_j\left[-\dfrac{1}{2k_{\mathrm{B}}T}\int^t_{-\infty}ds\lambda\dfrac{\partial}{\partial x^j_s}\left(V({\bm x}_s,X_s)-V({\bm x}_s,X_t)\right)\circ\dot{x}^j_s\right.\notag\\
&\quad\left.+\dfrac{1}{2\gamma k_{\mathrm{B}}T}\int^t_{-\infty}dsF^j({\bm x}_s)\lambda\dfrac{\partial}{\partial x^j_s}\left(V({\bm x}_s,X_s)-V({\bm x}_s,X_t)\right)\right.\notag\\
&\quad\left.-\dfrac{1}{4\gamma k_{\mathrm{B}}T}\int^t_{-\infty}ds\lambda^2\left(\left(\dfrac{\partial}{\partial x^j_s}V({\bm x}_s,X_s)\right)^2-\left(\dfrac{\partial}{\partial x^j_s}V({\bm x}_s,X_t)\right)^2\right)\right.\notag\\
&\quad\left.+\dfrac{1}{2\gamma}\int^t_{-\infty}ds\lambda\dfrac{\partial^2}{\partial(x^j_s)^2}\left(V({\bm x}_s,X_s)-V({\bm x}_s,X_t)\right)\right],
\end{align}
where the symbol $\circ$ denotes the multiplication in the sense of Stratonovich \cite{gardiner1985handbook}.
By using $V({\bm x}_s,X_s)=V({\bm x}_s,X_t)+(X_s-X_t)\partial_{X_t}V({\bm x}_s,X_t)+O((X_s-X_t)^2)$, we obtain to first order in $X_s-X_t$,
\begin{align}
-\mathcal{A}([{\bm x}]|[X])\simeq\dfrac{1}{2k_{\mathrm{B}}T}\left[\int^t_{-\infty}ds(X_s-X_t)\sum_j\dfrac{\partial}{\partial x^j_s}\Phi({\bm x}_s,X_t)\circ\dot{x}^j_s-\int^t_{-\infty}ds(X_s-X_t)\mathcal{L}^\dagger_s\Phi({\bm x}_s,X_t)\right],
\label{excess action}
\end{align}
where $\mathcal{L}^\dag_s$ denotes the backward operator (\ref{back ward operator}).
We remark that the first term on the right-hand side of (\ref{excess action}) is the entropic part while the second term is the so-called \textit{frenetic} part \cite{maes2020frenesy}.

We now decompose the interaction force $\Phi$ into a deterministic part $\langle \Phi|[X]\rangle$ and a fluctuating part $\eta_t:=\Phi-\langle \Phi|[X]\rangle$, where $\langle\cdot|[X]\rangle$ denotes the average with respect to $\mathbb{P}([{\bm x}]|[X])$:
\begin{align}
M\ddot{X}_t=\langle\Phi({\bm x}_t,X_t)|[X]\rangle-\Gamma\dot{X}_t+\sqrt{2\Gamma k_{\mathrm{B}}T}\Xi_t+\eta_t.
\end{align}
The deterministic part can be further decomposed into a streaming term and friction term as follows.
By using the relation (\ref{Girsanov}), we obtain
\begin{align}
\langle\Phi({\bm x}_t,X_t)|[X]\rangle-\langle\Phi({\bm x}_t,X_t)\rangle^{X_t}=-\langle\Phi({\bm x}_t,X_t);\mathcal{A}\rangle^{X_t}+O((X_s-X_t)^2).
\label{linear response}
\end{align}
Here, we have used $\langle\mathcal{A}\rangle^{X_t}=0$ to first order in $X_s-X_t$, which follows from the normalization condition $\langle\exp(-\mathcal{A})\rangle^{X_t}=1$.
By substituting (\ref{excess action}) into (\ref{linear response}), we obtain
\begin{align}
\langle \Phi({\bm x}_t,X_t)|[X]\rangle-\langle \Phi({\bm x}_t,X_t)\rangle^{X_t}=\int^t_{-\infty}ds(X_s-X_t)R_{\Phi\Phi}(t-s)+O((X_s-X_t)^2),
\label{linear response_R}
\end{align}
where
\begin{align}
R_{\Phi\Phi}(t-s):=\dfrac{1}{2k_{\mathrm{B}}T}\left[\dfrac{d}{ds}\langle \Phi({\bm x}_s,X_t);\Phi({\bm x}_t,X_t)\rangle^{X_t}-\langle\mathcal{L}^\dagger_s\Phi({\bm x}_s,X_t);\Phi({\bm x}_t,X_t)\rangle^{X_t}\right].
\end{align}
The function $R_{\Phi\Phi}(t-s)$ corresponds to the response function of the interaction force $\Phi$ against the perturbation of the probe position $X_s-X_t$:
\begin{equation}
R_{\Phi\Phi}(t-s)=\left.\dfrac{\delta \langle\Phi({\bm x}_t,X_t)|[X]\rangle}{\delta X_s}\right|_{X_s=X_t}.
\label{def: response function}
\end{equation}
We introduce $\gamma(t-s)$ via
\begin{align}
\gamma(t-s):=\int^s_{-\infty}du R_{\Phi\Phi}(t-u),\quad\text{for}\quad t\ge s.
\end{align}
Then, (\ref{linear response_R}) can be expressed as
\begin{align}
\langle \Phi({\bm x}_t,X_t)|[X]\rangle-\langle \Phi({\bm x}_t,X_t)\rangle^{X_t}=-\int^t_{-\infty}ds\gamma(t-s)\dot{X}_s+O((X_s-X_t)^2).
\end{align}
Thus, the deterministic part $\langle \Phi({\bm x}_t,X_t)|[X]\rangle$ is decomposed into the streaming term $G(X_t):=\langle \Phi({\bm x}_t,X_t)\rangle^{X_t}$ and friction term with the kernel $\gamma(t-s)$.

Next, we specify the statistical property of the fluctuating part $\eta_t$.
To leading order, it is given by
\begin{align}
\eta_t\simeq\eta^{(0)}_t:=\Phi({\bm x}_t,X_t)-\langle\Phi({\bm x}_t,X_t)\rangle^{X_t}.
\label{eta0}
\end{align}
The mean and two-time correlation of the noise $\eta^{(0)}_t$ read
\begin{align}
\langle\eta^{(0)}_t\rangle^{X_t}&=0,\\
\langle\eta^{(0)}_t\eta^{(0)}_s\rangle^{X_t}&=\langle \Phi({\bm x}_t,X_t);\Phi({\bm x}_s,X_t)\rangle^{X_t}.
\end{align}
Because the friction kernel $\gamma(t-s)$ can be expressed as
\begin{align}
\gamma(t-s)&=\dfrac{1}{k_{\mathrm{B}}T}\langle \Phi({\bm x}_t,X_t);\Phi({\bm x}_s,X_t)\rangle^{X_t}\notag\\
&\quad-\dfrac{1}{2k_{\mathrm{B}}T}\int^s_{-\infty}du\left[\dfrac{d}{du}\langle \Phi({\bm x}_u,X_t);\Phi({\bm x}_t,X_t)\rangle^{X_t}+\langle\mathcal{L}^\dagger_u\Phi({\bm x}_u,X_t);\Phi({\bm x}_t,X_t)\rangle^{X_t}\right],
\end{align}
the two-time correlation can be represented in terms of the friction kernel $\gamma(t-s)$ as
\begin{align}
\langle\eta^{(0)}_t\eta^{(0)}_s\rangle^{X_t}=k_{\mathrm{B}}T\left[\gamma(t-s)+\gamma_{\mathrm{ex}}(t-s)\right],\quad\text{for}\quad t\ge s,
\end{align}
where $\gamma_{\mathrm{ex}}(t-s)$ is the excess friction kernel defined by
\begin{align}
\gamma_{\mathrm{ex}}(t-s):=\dfrac{1}{2k_{\mathrm{B}}T}\int^s_{-\infty}du\left[\dfrac{d}{du}\langle \Phi({\bm x}_u,X_t);\Phi({\bm x}_t,X_t)\rangle^{X_t}+\langle\mathcal{L}^\dagger_u\Phi({\bm x}_u,X_t);\Phi({\bm x}_t,X_t)\rangle^{X_t}\right].
\end{align}

To summarize, the effective dynamics of the probe is given by
\begin{align}
M\ddot{X}_t=G(X_t)-\int^t_{-\infty}ds\gamma(t-s)\dot{X}_s-\Gamma\dot{X}_t+\sqrt{2\Gamma k_{\mathrm{B}}T}\Xi_t+\eta_t,
\end{align}
where 
\begin{align}
G(X_t):=\langle \Phi({\bm x}_t,X_t)\rangle^{X_t},
\end{align}
\begin{align}
\gamma(t-s):=\dfrac{1}{2k_{\mathrm{B}}T}\int^s_{-\infty}du\left[\dfrac{d}{du}\langle \Phi({\bm x}_u,X_t);\Phi({\bm x}_t,X_t)\rangle^{X_t}-\langle\mathcal{L}^\dagger_u\Phi({\bm x}_u,X_t);\Phi({\bm x}_t,X_t)\rangle^{X_t}\right],\quad\text{for}\quad t\ge s,
\label{derivation: friction kerel}
\end{align}
and we have rewritten $\eta^{(0)}_t$ as $\eta_t$ for notational simplicity.
The noise intensity is related to the friction kernel in the following form:
\begin{align}
\langle\eta_t\eta_s\rangle^{X_t}=k_{\mathrm{B}}T\left[\gamma(t-s)+\gamma_{\mathrm{ex}}(t-s)\right],\quad\text{for}\quad t\ge s,
\end{align}
with the excess friction kernel $\gamma_{\mathrm{ex}}(t-s)$ given by
\begin{align}
\gamma_{\mathrm{ex}}(t-s):=\dfrac{1}{2k_{\mathrm{B}}T}\int^s_{-\infty}du\left[\dfrac{d}{du}\langle \Phi({\bm x}_u,X_t);\Phi({\bm x}_t,X_t)\rangle^{X_t}+\langle\mathcal{L}^\dagger_u\Phi({\bm x}_u,X_t);\Phi({\bm x}_t,X_t)\rangle^{X_t}\right].
\end{align}

\subsection{Derivation of the inequality that bounds the violation of the second FDR}
Here, we derive our first main result (\ref{bound on the violation of the 2nd FDR}).
We use the fact that the violation of the second FDR is originated from that of the first FDR.
The excess friction kernel $\gamma_{\mathrm{ex}}(t-s)$ can be expressed as
\begin{align}
\gamma_{\mathrm{ex}}(t-s)&=\dfrac{1}{k_{\mathrm{B}}T}\langle\eta_t\eta_s\rangle^{X_t}-\gamma(t-s)\notag\\
&=\int^s_{-\infty}du\left[\dfrac{1}{k_{\mathrm{B}}T}\partial_u C_{\Phi\Phi}(t-u)-R_{\Phi\Phi}(t-u)\right],
\label{gamma_ex in terms of first FDR}
\end{align}
where $C_{\Phi\Phi}(t-s)$ denotes the connected correlation function,
\begin{equation}
C_{\Phi\Phi}(t-s):=\langle \Phi({\bm x}_t,X_t);\Phi({\bm x}_s,X_t)\rangle^{X_t}.
\end{equation}
While the standard first FDR holds in equilibrium, $k_{\mathrm{B}}TR_{\Phi\Phi}(t-s)=\partial_sC_{\Phi\Phi}(t-s)$ \cite{kubo2012statistical,risken1996fokker}, the following Seifert-Speck generalized FDR holds in the nonequilibrium steady state \cite{seifert2010fluctuation}:
\begin{equation}
R_{\Phi\Phi}(t-s)-\dfrac{1}{k_{\mathrm{B}}T}\partial_sC_{\Phi\Phi}(t-s)=-\partial_s\left\langle\Phi({\bm x}_t,X_t)\dfrac{\partial}{\partial X_t}\Delta s^{X_t}_{\mathrm{tot}}\right\rangle^{X_t},
\label{Seifert-Speck GFDR}
\end{equation}
where $\Delta s^{X_t}_{\mathrm{tot}}$ denotes the total stochastic entropy production up to time $s$ of the nonequilibrium medium in the NESS with $X_t$ held fixed \cite{seifert2005entropy,sekimoto2010stochastic,seifert2012stochastic}:
\begin{align}
\Delta s^{X_t}_{\mathrm{tot}}=-\ln P^{X_t}_{\mathrm{ss}}({\bm x}_u)\Bigl|^s_{-\infty}+\int^s_{-\infty}du\dfrac{1}{k_{\mathrm{B}}T}\sum_j\left[F^j({\bm x}_u)-\lambda\dfrac{\partial}{\partial x^j_u}V({\bm x}_u,X_t)\right]\circ\dot{x}^j_u.
\end{align}
Here, the first two terms represent the stochastic Shannon entropy difference, while the last term represents the stochastic entropy production of the equilibrium thermal bath.
From this expression, $\partial_{X_t}\Delta s^{X_t}_{\mathrm{tot}}$ reads
\begin{align}
\dfrac{\partial}{\partial X_t}\Delta s^{X_t}_{\mathrm{tot}}&=-\dfrac{\partial}{\partial X_t}\ln P^{X_t}_{\mathrm{ss}}({\bm x}_u)\biggl|^s_{-\infty}+\int^s_{-\infty}du\dfrac{1}{k_{\mathrm{B}}T}\sum_j\dfrac{\partial}{\partial x^j_u}\Phi({\bm x}_u,X_t)\circ\dot{x}^j_u\notag\\
&=\left(-\dfrac{\partial}{\partial X_t}\ln P^{X_t}_{\mathrm{ss}}({\bm x}_u)+\dfrac{1}{k_{\mathrm{B}}T}\Phi({\bm x}_u,X_t)\right)\biggl|^s_{-\infty}.
\label{entropy_reponse}
\end{align}
Hence, $\langle\partial_{X_t}\Delta s^{X_t}_{\mathrm{tot}}\rangle^{X_t}=0$.
By substituting (\ref{Seifert-Speck GFDR}) into (\ref{gamma_ex in terms of first FDR}) and using the Cauchy-Schwarz inequality, we thus obtain
\begin{align}
\left|\gamma_{\mathrm{ex}}(t-s)\right|&=\left|\int^s_{-\infty}du\partial_u\left\langle\Phi({\bm x}_t,X_t)\dfrac{\partial}{\partial X_t}\Delta s^{X_t}_{\mathrm{tot}}\right\rangle^{X_t}\right|\notag\\
&=\left|\left\langle\Phi({\bm x}_t,X_t)\dfrac{\partial}{\partial X_t}\Delta s^{X_t}_{\mathrm{tot}}\right\rangle^{X_t}\right|\notag\\
&\le\sqrt{\left\langle(\Phi({\bm x}_t,X_t))^2\right\rangle^{X_t}}\sqrt{\mathrm{Var}\left[\partial_{X_t}\Delta s^{X_t}_{\mathrm{tot}}\right]}.
\label{derivation: bound}
\end{align}

\section{Validity of the effective dynamics\label{Validity of the effective dynamics}}
In this section, we discuss the validity of the effective probe dynamics (\ref{Effective generalized Langevin}), which is derived by using nonequilibrium linear response theory.
There are mainly two subtle points in the derivation.
First, it is not clear whether the noise term $\eta_t=\Phi({\bm x}_t,X_t)-\langle\Phi({\bm x}_t,X_t)\rangle^{X_t}$ and friction kernel $\gamma(t-s)$ exclude the slow modes associated with the probe motion \cite{zwanzig1972memory,sture1974strategies,mori1973nonlinear,kawasaki1973simple,fujisaka1976fluctuation}.
It is also unclear whether the noise and friction kernel include the effects of hydrodynamic fields \cite{granek2021anomalous}, the properties of which have been well investigated in equilibrium systems \cite{alder1970decay,pomeau1975time,huang2011direct,franosch2011resonances,kheifets2014observation}.
In this regard, we conjecture that nonequilibrium linear response theory can describe such hydrodynamic effects because the response function $R_{\Phi\Phi}$ may exhibits a long time tail if the back reaction of the particles against the probe motion propagates with slow time scales.
Second, the condition for the time-scale separation between the probe and the particles is ambiguous.
Specifically, the validity of the expansion of the excess action $\mathcal{A}([{\bm x}]|[X])$ in terms of $X_s-X_t$, (\ref{excess action}), is unclear because the excess action includes the integral from time $t=-\infty$.

The guiding principle here is that if a system allows a phenomenological description at the mesoscale, then it should be uniquely determined.
We thus aim to verify the validity of the effective probe dynamics (\ref{Effective generalized Langevin}) (or (\ref{Effective Langevin})) by deriving it through other methods.
One of the most frequently used methods is the projection operator method \cite{zwanzig1961memory}.
This method is based on the natural idea of singling out slow degrees of freedom and has recently been applied even to nonlinear lattices to derive fluctuating hydrodynamics \cite{saito2021microscopic}.
Because various representations can be obtained depending on the definitions of the projection operator, the most crucial point in this approach is to clarify which definition is consistent with the description at the mesoscale.
However, it is generally difficult to provide the condition that uniquely characterizes the projection operator corresponding to the mesoscopic description.
In this regard, adiabatic perturbation theory is a systematic method based on more explicit assumptions \cite{d2016negative,weinberg2017adiabatic,granek2021anomalous}.
In this theory, the probability distribution of the fast variables is expanded around a steady state in terms of the time derivative of the slow variables.
A derivation based on the linearized Dean equation has also been recently proposed \cite{feng2021effective}.

While all of these methods are expected to provide results consistent with nonequilibrium linear response theory, here we use the singular perturbation method developed in \cite{nakayama2018unattainability}.
This approach is based on a clear assumption about the separation of time scales and thus allows us to derive slow dynamics systematically.
In the following, we explain the details of the derivation of the effective dynamics using the singular perturbation method and show that the resulting effective dynamics corresponds to (\ref{Effective Langevin}).

\subsection{Singular perturbation method}
We first rewrite the model (\ref{Model_probe}) and (\ref{Model_particles}) in the following form:
\begin{align}
\dot{X}_t&=\dfrac{P_t}{M},\\
\dot{P}_t&=\Phi({\bm x}_t,X_t)-\Gamma\dfrac{P_t}{M}+\sqrt{2\Gamma k_{\mathrm{B}}T}\Xi_t,\\
\dot{x}^j_t&=\dfrac{1}{\gamma}\left[F^j({\bm x}_t)-\lambda\dfrac{\partial}{\partial x^j_t}V({\bm x}_t,X_t)\right]+\sqrt{\dfrac{2k_{\mathrm{B}}T}{\gamma}}\xi^j_t,
\end{align}
where $P_t$ denotes the momentum of the probe.
The corresponding Fokker-Planck equation for the probability density $\rho_t(X,P,{\bm x})$ reads
\begin{align}
\dfrac{\partial}{\partial t}\rho_t&=-\dfrac{P}{M}\dfrac{\partial}{\partial X}\rho_t+\dfrac{\partial}{\partial P}\left[\left(-\Phi({\bm x},X)+\dfrac{\Gamma}{M}P\right)\rho_t\right]+\Gamma k_{\mathrm{B}}T\dfrac{\partial^2}{\partial P^2}\rho_t\notag\\
&\quad+\sum_j\left[\dfrac{1}{\gamma}\dfrac{\partial}{\partial x^j}\left[\left(-F^j({\bm x})+\lambda\dfrac{\partial}{\partial x^j}V({\bm x},X)\right)\rho_t\right]+\dfrac{k_{\mathrm{B}}T}{\gamma}\dfrac{\partial^2}{\partial (x^j)^2}\rho_t\right].
\label{FP}
\end{align}

We are now interested in the regime where the motion of the probe is much slower than that of the particles.
To describe this regime more explicitly, we first explain several characteristic time scales for this system.
Let $\ell$ be the characteristic length scale associated with the coupling potential $V$.
Here, we have in mind a case where $V$ is a confining potential.
Otherwise, hydrodynamic modes may emerge, of which time scales comparable to the time scales of the probe.
The probe has two time scales: the characteristic time scale for the probe to relax in the coupling potential, $\tau_X:=\sqrt{M\ell^2/k_{\mathrm{B}}T}$, and the momentum relaxation time, $\tau_P:=M/\Gamma$.
Similarly, the particles diffuse in the coupling potential with the time scale $\tau_c:=\gamma\ell^2/k_{\mathrm{B}}T$.
We denote by $\tau_m$ the time scale for the particles to relax to the steady state.
Specifically, $\tau_m$ is defined through the spectral gap $\Delta_m$ between the first two largest eigenvalues of the Fokker-Planck operator for the particles (see (\ref{L_m}), below).
That is, if we denote by $\Lambda_0$ and $\Lambda_1$ the first two largest eigenvalues, the spectral gap is given by $\Delta_m:=-\mathrm{Re}[\Lambda_1]>0$, because $\Lambda_0=0$ and $\mathrm{Re}[\Lambda_1]<0$ from the Perron-Frobenius theorem.
Note that $\tau_m$ may correspond to the time scale associated with the hydrodynamic modes, especially when the coupling potential $V$ is not a confining potential.
The most crucial assumption in the singular perturbation method is the separation of time scales:
\begin{equation}
\tau_P\sim\tau_X\gg\tau_c\sim\tau_m.
\label{time scale separation}
\end{equation}
This condition implies that there are no slow modes associated with the particles that are comparable to the motion of the probe.
Hereafter, we consider the dynamics on the fast time scale $\tau:=t/\tau_c$.

To identify small parameters in (\ref{FP}), we introduce dimensionless variables.
We define $\tilde{X}:=X/\ell$, $\tilde{x}^j:=x^j/\ell$, and $\tilde{P}:=P/\sqrt{Mk_{\mathrm{B}}T}$.
We also define the dimensionless potential and force as $\tilde{\Phi}(\tilde{\bm x},\tilde{X}):=-\lambda\partial\tilde{V}(\tilde{\bm x},\tilde{X})/\partial\tilde{X}$ with $\tilde{V}(\tilde{\bm x},\tilde{X}):=V({\bm x},X)/k_{\mathrm{B}}T$ and $\tilde{F}(\tilde{\bm x}):=F({\bm x})\ell/k_{\mathrm{B}}T$. 
Correspondingly, we write the probability density as $\tilde{\rho}_\tau(\tilde{X},\tilde{P},\tilde{\bm x}):=\rho_{\tau_c\tau}(\ell\tilde{X},\sqrt{Mk_{\mathrm{B}}T}\tilde{P},\ell\tilde{\bm x})$.
Then, (\ref{FP}) can be rewritten as
\begin{align}
\dfrac{\partial}{\partial\tau}\tilde{\rho}_\tau&=-\dfrac{\tau_c}{\tau_X}\tilde{P}\dfrac{\partial}{\partial\tilde{X}}\tilde{\rho}_\tau+\dfrac{\tau_c}{\tau_X}\dfrac{\partial}{\partial\tilde{P}}\left[-\tilde{\Phi}(\tilde{\bm x},\tilde{X})\tilde{\rho}_\tau\right]+\dfrac{\tau_c}{\tau_P}\dfrac{\partial}{\partial\tilde{P}}\left(\tilde{P}\tilde{\rho}_\tau\right)+\dfrac{\tau_c}{\tau_P}\dfrac{\partial^2}{\partial\tilde{P}^2}\tilde{\rho}_\tau\notag\\
&\quad+\sum_j\left[\dfrac{\partial}{\partial\tilde{x}^j}\left[\left(-\tilde{F}(\tilde{\bm x})+\dfrac{\partial}{\partial\tilde{x}^j}\tilde{V}(\tilde{\bm x},\tilde{X})\right)\tilde{\rho}_\tau\right]+\dfrac{\partial^2}{\partial(\tilde{x}^j)^2}\tilde{\rho}_\tau\right].
\label{FP-dimensionless}
\end{align}
From the condition (\ref{time scale separation}), (\ref{FP-dimensionless}) can be expressed in terms of the small parameter $\epsilon:=\tau_c/\tau_X\ll1$ as
\begin{align}
\dfrac{\partial}{\partial\tau}\tilde{\rho}_\tau=(\epsilon\mathcal{L}_{\mathrm{pb}}+\mathcal{L}_m)\tilde{\rho}_\tau,
\label{FP-dimensionless-epsilon}
\end{align}
where $\mathcal{L}_{\mathrm{pb}}$ and $\mathcal{L}_m$ denote the Fokker-Planck operators for the probe and the nonequilibrium medium, respectively:
\begin{align}
\mathcal{L}_{\mathrm{pb}}:=-\tilde{P}\dfrac{\partial}{\partial\tilde{X}}-\dfrac{\partial}{\partial\tilde{P}}\tilde{\Phi}(\tilde{\bm x},\tilde{X})+\dfrac{\tau_X}{\tau_P}\dfrac{\partial}{\partial\tilde{P}}\tilde{P}+\dfrac{\tau_X}{\tau_P}\dfrac{\partial^2}{\partial\tilde{P}^2},
\end{align}
\begin{align}
\mathcal{L}_m:=\sum_j\left[\dfrac{\partial}{\partial\tilde{x}^j}\left(-\tilde{F}(\tilde{\bm x})+\dfrac{\partial}{\partial\tilde{x}^j}\tilde{V}(\tilde{\bm x},\tilde{X})\right)+\dfrac{\partial^2}{\partial(\tilde{x}^j)^2}\right].
\label{L_m}
\end{align}
The form of (\ref{FP-dimensionless-epsilon}) implies that the system first relaxes toward the slow manifold characterized by $\mathcal{L}_m$ on the fast time scale $\tau\sim1$ and then evolves slowly on the slow manifold.
The motion on the slow manifold is characterized by the following equation for the reduced probability density $R_\tau(\tilde{X},\tilde{P}):=\int \prod_jd\tilde{x}^j\tilde{\rho}_\tau(\tilde{X},\tilde{P},\tilde{\bm x})$, which is obtained by integrating out $\tilde{\bm x}$ in (\ref{FP-dimensionless-epsilon}):
\begin{align}
\dfrac{\partial}{\partial\tau}R_\tau=\epsilon\left[-\tilde{P}\dfrac{\partial}{\partial\tilde{X}}R_\tau-\dfrac{\partial}{\partial\tilde{P}}\int\prod_jd\tilde{x}^j\tilde{\Phi}(\tilde{\bm x},\tilde{X})\tilde{\rho}_\tau+\dfrac{\tau_X}{\tau_P}\dfrac{\partial}{\partial\tilde{P}}(\tilde{P}R_\tau)+\dfrac{\tau_X}{\tau_P}\dfrac{\partial^2}{\partial\tilde{P}^2}R_\tau\right].
\end{align}
Because $R_\tau$ evolves slowly, secular terms arise in the naive perturbation expansion $\tilde{\rho}_\tau=\tilde{\rho}^{(0)}_\tau+\epsilon\tilde{\rho}^{(1)}_\tau+\cdots$.
Therefore, to describe the dynamics on the slow manifold, we assume that the $\tau$-dependence of $\tilde{\rho}_\tau$ is expressed in terms of the $\tau$-dependent operator $M_\tau$ that acts on the reduced probability density $R_\tau$:
\begin{align}
\tilde{\rho}_\tau(\tilde{X},\tilde{P},\tilde{\bm x})=M_\tau[R_\tau](\tilde{X},\tilde{P},\tilde{\bm x}).
\end{align} 
From this functional ansatz, we can decompose the $\tau$-dependence of $\tilde{\rho}_\tau$ into its explicit and implicit parts through $R_\tau$.
Correspondingly, we introduce $\Omega_\tau$ as the $\tau$-dependent operator that represents the slow dynamics:
\begin{align}
\Omega_\tau[R_\tau](\tilde{X},\tilde{P}):=\epsilon\left[-\tilde{P}\dfrac{\partial}{\partial\tilde{X}}R_\tau-\dfrac{\partial}{\partial\tilde{P}}\int\prod_jd\tilde{x}^j\tilde{\Phi}(\tilde{\bm x},\tilde{X})M_\tau[R_\tau]+\dfrac{\tau_X}{\tau_P}\dfrac{\partial}{\partial\tilde{P}}(\tilde{P}R_\tau)+\dfrac{\tau_X}{\tau_P}\dfrac{\partial^2}{\partial\tilde{P}^2}R_\tau\right].
\label{FP-R}
\end{align}
In terms of $M_\tau$ and $\Omega_\tau$, (\ref{FP-dimensionless-epsilon}) can be expressed as
\begin{align}
\dfrac{\partial}{\partial\tau}M_\tau[R_\tau]+\int\dfrac{\delta M_\tau[R_\tau]}{\delta R_\tau}\Omega_\tau[R_\tau]d\tilde{X}d\tilde{P}=(\epsilon\mathcal{L}_{\mathrm{pb}}+\mathcal{L}_m)M_\tau[R_\tau].
\label{FP-dimensionless-epsilon-M}
\end{align}

We now assume that $M_\tau$ and $\Omega_\tau$ have asymptotic expansions in terms of the asymptotic sequences $\{\epsilon^n\}^\infty_{n=0}$ as $\epsilon\rightarrow0$:
\begin{align}
M_\tau&=M^{(0)}_\tau+\epsilon M^{(1)}_\tau+\epsilon^2M^{(2)}_\tau+\cdots,\\
\Omega_\tau&=\epsilon\Omega^{(1)}_\tau+\epsilon^2\Omega^{(2)}_\tau+\cdots.
\end{align}
Note that $\Omega^{(0)}_\tau$ is set to zero because of the form (\ref{FP-R}).
The leading order of (\ref{FP-dimensionless-epsilon-M}) gives
\begin{align}
\dfrac{\partial}{\partial\tau}M^{(0)}_\tau[R_\tau]=\mathcal{L}_mM^{(0)}_\tau[R_\tau].
\end{align}
From this equation, it follows that
\begin{align}
M^{(0)}_\tau[R_\tau](\tilde{X},\tilde{P},\tilde{\bm x})\simeq R_\tau(\tilde{X},\tilde{P})Q_{\mathrm{ss}}(\tilde{\bm x}|\tilde{X})
\label{M_0}
\end{align}
for $\tau\gg1$, where $Q_{\mathrm{ss}}(\tilde{\bm x}|\tilde{X})$ denotes the stationary distribution for $\tilde{\bm x}$ under the condition that $\tilde{X}$ is held fixed:
\begin{equation}
\mathcal{L}_mQ_{\mathrm{ss}}=0.
\end{equation}
Here, we have imposed the condition 
\begin{equation}
R_\tau=\int\prod_jd\tilde{x}^jM^{(0)}_\tau[R_\tau].
\end{equation}
Note that, in the approximation in (\ref{M_0}), the additional terms are ignored because they decay exponentially with the time scale of $\tau_c$.
By substituting (\ref{M_0}) into (\ref{FP-R}), we obtain
\begin{align}
\Omega^{(1)}_\tau[R_\tau]\simeq-\tilde{P}\dfrac{\partial}{\partial\tilde{X}}R_\tau-\dfrac{\partial}{\partial\tilde{P}}\left[\langle\tilde{\Phi}(\tilde{\bm x},\tilde{X})\rangle^{X}R_\tau\right]+\dfrac{\tau_X}{\tau_P}\dfrac{\partial}{\partial\tilde{P}}(\tilde{P}R_\tau)+\dfrac{\tau_X}{\tau_P}\dfrac{\partial^2}{\partial\tilde{P}^2}R_\tau
\label{Omega_1}
\end{align}
for $\tau\gg1$, where $\langle\cdot\rangle^{X}$ denotes the average with respect to $Q_{\mathrm{ss}}$.
The subleading order of (\ref{FP-dimensionless-epsilon-M}) gives 
\begin{align}
\dfrac{\partial}{\partial\tau}M^{(1)}_\tau[R_\tau]+\int\dfrac{\delta M^{(0)}_\tau[R_\tau]}{\delta R_\tau}\Omega^{(1)}_\tau[R_\tau]d\tilde{X}d\tilde{P}=\mathcal{L}_{\mathrm{pb}}M^{(0)}_\tau[R_\tau]+\mathcal{L}_mM^{(1)}_\tau[R_\tau].
\label{FP-dimensionless-epsilon-M_1}
\end{align}
For $\tau\gg1$, we obtain the first-order solution $M^{(1)}_\tau$ as
\begin{align}
M^{(1)}_\tau[R_\tau]&\simeq\mathcal{L}^{-1}_m\left[Q_{\mathrm{ss}}\Omega^{(1)}_\tau[R_\tau]-\mathcal{L}_{\mathrm{pb}}M^{(0)}_\tau[R_\tau]\right]\notag\\
&=-\int^\infty_0dse^{s\mathcal{L}_m}\left\{Q_{\mathrm{ss}}\dfrac{\partial}{\partial\tilde{P}}\left[\left(\tilde{\Phi}(\tilde{\bm x},\tilde{X})-\langle\tilde{\Phi}(\tilde{\bm x},\tilde{X})\rangle^{X}\right)R_\tau\right]+\tilde{P}R_\tau\dfrac{\partial}{\partial\tilde{X}}Q_{\mathrm{ss}}\right\}\notag\\
&=-\left(\dfrac{\partial}{\partial\tilde{P}}R_\tau\right)\int^\infty_0dse^{s\mathcal{L}_m}\left[Q_{\mathrm{ss}}\left(\tilde{\Phi}(\tilde{\bm x},\tilde{X})-\langle\tilde{\Phi}(\tilde{\bm x},\tilde{X})\rangle^{X}\right)\right]-\tilde{P}R_\tau\int^\infty_0dse^{s\mathcal{L}_m}\dfrac{\partial}{\partial\tilde{X}}Q_{\mathrm{ss}}.
\label{M_1}
\end{align}
We note that in order for the above expression for $M^{(1)}_\tau$ to be well-defined, it is necessary that $Q_{\mathrm{ss}}\Omega^{(1)}_\tau[R_\tau]-\mathcal{L}_{\mathrm{pb}}M^{(0)}_\tau[R_\tau]$ does not include the zero eigenfunction of $\mathcal{L}_m$.
This solvability condition is nothing but (\ref{Omega_1}).
By substituting (\ref{M_1}) into (\ref{FP-R}), we obtain
\begin{align}
\Omega^{(2)}_\tau[R_\tau]&=-\dfrac{\partial}{\partial\tilde{P}}\int\prod_jd\tilde{x}^j\tilde{\Phi}(\tilde{\bm x},\tilde{X})M^{(1)}_\tau[R_\tau]\notag\\
&=\int^\infty_0ds\langle\tilde{\Phi}(\tilde{\bm x}_s,\tilde{X});\tilde{\Phi}(\tilde{\bm x}_0,\tilde{X})\rangle^{X}\dfrac{\partial^2}{\partial\tilde{P}^2}R_\tau+\dfrac{\partial}{\partial\tilde{P}}\left(\int^\infty_0ds\left\langle\tilde{\Phi}(\tilde{\bm x}_s,\tilde{X})\dfrac{\partial}{\partial\tilde{X}}\ln Q_{\mathrm{ss}}(\tilde{\bm x}_0|\tilde{X})\right\rangle^{X}\tilde{P}R_\tau\right).
\label{Omega_2}
\end{align}
From (\ref{FP-R}), (\ref{Omega_1}), and (\ref{Omega_2}), the effective dynamics for the slow variable $R_\tau$ is given by
\begin{align}
\dfrac{\partial}{\partial\tau}R_\tau&\simeq\epsilon\left[-\tilde{P}\dfrac{\partial}{\partial\tilde{X}}R_\tau-\dfrac{\partial}{\partial\tilde{P}}\left[\langle\tilde{\Phi}(\tilde{\bm x},\tilde{X})\rangle^{X}R_\tau\right]+\dfrac{\tau_X}{\tau_P}\dfrac{\partial}{\partial\tilde{P}}(\tilde{P}R_\tau)+\dfrac{\tau_X}{\tau_P}\dfrac{\partial^2}{\partial\tilde{P}^2}R_\tau\right]\notag\\
&\quad+\epsilon^2\left[\int^\infty_0ds\langle\tilde{\Phi}(\tilde{\bm x}_s,\tilde{X});\tilde{\Phi}(\tilde{\bm x}_0,\tilde{X})\rangle^{X}\dfrac{\partial^2}{\partial\tilde{P}^2}R_\tau+\dfrac{\partial}{\partial\tilde{P}}\left(\int^\infty_0ds\left\langle\tilde{\Phi}(\tilde{\bm x}_s,\tilde{X})\dfrac{\partial}{\partial\tilde{X}}\ln Q_{\mathrm{ss}}(\tilde{\bm x}_0|\tilde{X})\right\rangle^{X}\tilde{P}R_\tau\right)\right]\notag\\
&=\tau_c\left[-\dfrac{P}{M}\dfrac{\partial}{\partial X}R_\tau-\dfrac{\partial}{\partial P}\left[\left(\langle\Phi(\bm x_s,X)\rangle^{X}-\dfrac{\Gamma}{M}P\right)R_\tau\right]+\Gamma k_{\mathrm{B}}T\dfrac{\partial^2}{\partial P^2}R_\tau\right]\notag\\
&\quad+\tau_c\left[\int^\infty_0dt\langle\Phi({\bm x}_t,X);\Phi({\bm x}_0,X)\rangle^{X}\dfrac{\partial^2}{\partial P^2}R_\tau+\dfrac{\partial}{\partial P}\left(\int^\infty_0dt\left\langle\Phi({\bm x}_t,X)\dfrac{\partial}{\partial X}\ln P^{X}_{\mathrm{ss}}({\bm x}_0)\right\rangle^{X}\dfrac{P}{M}R_\tau\right)\right],
\end{align}
where $P^{X}_{\mathrm{ss}}({\bm x}):=Q_{\mathrm{ss}}(\ell^{-1}{\bm x}|\ell^{-1}X)$.
Therefore, the effective Langevin equation for the probe reads
\begin{align}
\dot{X}_t&=\dfrac{P_t}{M},\label{singular_perturbation: result 1}\\
\dot{P}_t&=G(X_t)-(\Gamma+\gamma_{\mathrm{eff}})\dfrac{P_t}{M}+\sqrt{2(\Gamma+\gamma_{\mathrm{eff}}+\gamma_{\mathrm{ex}})k_{\mathrm{B}}T}\Xi_t.\label{singular_perturbation: result 2}
\end{align}
Here, $G(X_t)$ denotes the streaming term,
\begin{align}
G(X_t):=\left\langle\Phi({\bm x}_t,X_t)\right\rangle^{X_t},
\end{align}
and $\gamma_{\mathrm{eff}}$ denotes the effective friction coefficient,
\begin{align}
\gamma_{\mathrm{eff}}&:=\int^\infty_0ds\left\langle\Phi({\bm x}_s,X_t)\dfrac{\partial}{\partial X_t}\ln P^{X_t}_{\mathrm{ss}}({\bm x}_0)\right\rangle^{X_t}.
\end{align}
Note that the integrand in the above expression can be expressed in terms of the response function $R_{\Phi\Phi}(t-u)$, (\ref{def: response function}), by using the Seifert-Speck generalized FDR (\ref{Seifert-Speck GFDR}) \cite{seifert2010fluctuation}:
\begin{align}
\left\langle\Phi({\bm x}_s,X_t)\dfrac{\partial}{\partial X_t}\ln P^{X_t}_{\mathrm{ss}}({\bm x}_0)\right\rangle^{X_t}=\int^0_{-\infty}duR_{\Phi\Phi}(s-u),
\end{align}
where
\begin{align}
R_{\Phi\Phi}(s-u)&=\dfrac{\partial}{\partial u}\left\langle\Phi({\bm x}_s,X_t)\dfrac{\partial}{\partial X_t}\ln P^{X_t}_{\mathrm{ss}}({\bm x}_u)\right\rangle^{X_t}.
\end{align}
Finally, $\gamma_{\mathrm{ex}}$ denotes the excess friction coefficient,
\begin{align}
\gamma_{\mathrm{ex}}&:=\dfrac{1}{k_{\mathrm{B}}T}\int^\infty_0ds\langle\Phi({\bm x}_s,X_t);\Phi({\bm x}_0,X_t)\rangle^{X_t}-\gamma_{\mathrm{eff}}\notag\\
&=\dfrac{1}{k_{\mathrm{B}}T}\int^\infty_0ds\langle\eta^{(0)}_s\eta^{(0)}_0\rangle^{X_t}-\gamma_{\mathrm{eff}},
\end{align}
where $\eta^{(0)}_t$ is defined by (\ref{eta0}).
Therefore, (\ref{singular_perturbation: result 1}) and (\ref{singular_perturbation: result 2}) exactly correspond to (\ref{Effective Langevin}), and thus the singular perturbation method and nonequilibrium linear response theory give the same result.

\section{Example: Potential switching medium\label{Example}}
We here present a simple model for a nonequilibrium medium as an example of the previous results.
In this model, the particles are driven by potentials that switch stochastically.
We can confirm that the effective dynamics is consistent with the exact solution because all relevant quantities can be calculated explicitly.
Furthermore, in the fast switching limit, this model provides an example of a nonequilibrium medium where the second FDR holds.
We can show that the upper bound of the inequality (\ref{bound on the violation of the 2nd FDR}) goes to zero in this limit.

\subsection{Model}
The time evolution of $X_t$ is given by the following underdamped Langevin equation:
\begin{align}
M\ddot{X}_t=\Phi({\bm x}_t,X_t)-\Gamma\dot{X}_t+\sqrt{2\Gamma k_{\mathrm{B}}T}\Xi_t
\label{ex:Model_probe}
\end{align}
with $\Phi({\bm x}_t,X_t):=-\lambda\partial V({\bm x}_t,X_t)/\partial X_t$.
We suppose that the probe is linearly coupled to the particles.
That is, $V({\bm x},X)$ is a harmonic potential with the spring constant $\kappa_c$:
\begin{align}
\Phi({\bm x}_t,X_t)=-\lambda\kappa_c\sum_j(X_t-x^j_t).
\end{align}

The particles are described by the so-called \textit{potential switching model}, i.e., they are subjected to potentials that switch stochastically \cite{wang2016entropy,dieterich2015single,toyabe2007experimental}:
\begin{align}
\gamma\dot{x}^j_t=-\kappa_b(x^j_t-\sigma^j_tL)-\lambda\kappa_c(x^j_t-X_t)+\sqrt{2\gamma k_{\mathrm{B}}T}\xi^j_t.
\label{ex:Model_particles}
\end{align}
The first term on the right-hand side of (\ref{ex:Model_particles}) denotes the force induced by the switching potential with the spring constant $\kappa_b$ and switching width $L$.
Here, $\sigma^j_t\in\{0,1\}$ denotes the potential state of the $j$-th particle at time $t$, which switches stochastically between 0 and 1 at a rate $r$ independently for each particle.
We note that this model is an example of (\ref{Model_particles}) with $F^j({\bm x})=-\kappa_b(x^j-\sigma^jL)$ and $V({\bm x},X)=\sum_j\kappa_c(x^j-X)^2/2$.
Since the transition rates are equal for the transitions from 0 to 1 and 1 to 0, there is no entropy production associated with these transitions.
Therefore, this model satisfies the LDB for the combined set of variables $(x^j,\sigma^j)$.
We remark that this model can also be regarded as a run-and-tumble model \cite{tailleur2008statistical}.

\subsection{Effective dynamics}
Even in this model, we can derive the effective dynamics of the probe by using nonequilibrium linear response theory (see Appendix \ref{Explicit calculation} for the detailed derivation).
The effective dynamics of the probe is given by the following generalized Langevin-type equation:
\begin{align}
M\ddot{X}_t=G(X_t)-\Gamma\dot{X}_t-\int^t_{-\infty}ds\gamma(t-s)\dot{X}_s+\sqrt{2\Gamma k_{\mathrm{B}}T}\Xi_t+\eta_t.
\label{ex:Effective generalized Langevin}
\end{align}
The streaming term $G(X_t)\:=\langle\Phi({\bm x}_t,X_t)\rangle^{X_t}$ is expressed as
\begin{align}
G(X_t)=-\dfrac{N\lambda\kappa_c\kappa_b}{\kappa_b+\lambda\kappa_c}\left(X_t-\dfrac{L}{2}\right),
\end{align}
where $\langle\cdot\rangle^{X_t}$ denotes the average with respect to the stationary distribution $P^{X_t}_{\mathrm{ss}}({\bm x},{\bm\sigma})$ for the particle dynamics (\ref{ex:Model_particles}) with $X_t$ held fixed, where ${\bm \sigma}:=\{\sigma^1,\sigma^2,\cdots,\sigma^N\}$.
The friction kernel is given by
\begin{align}
\gamma(t-s)=\dfrac{N\lambda^2\kappa^2_c}{\kappa_b+\lambda\kappa_c}e^{-\frac{t-s}{\tau_x}},\quad\text{for}\quad t\ge s,
\label{friction kernel}
\end{align}
where $\tau_x:=\gamma/(\kappa_b+\lambda\kappa_c)$ denotes the characteristic time scale for the particles to relax in the coupling and switching potentials.
The expression (\ref{friction kernel}) states that dissipation happens on the time scale $\tau_x$.
By contrast, the noise correlation $\langle\eta_t\eta_s\rangle^{X_t}$ additionally includes the switching time scale $\tau_r:=1/2r$:
\begin{align}
\langle\eta_t\eta_s\rangle^{X_t}=k_{\mathrm{B}}T\left[\gamma(t-s)+\gamma_{\mathrm{ex}}(t-s)\right],
\end{align}
where $\gamma_{\mathrm{ex}}(t-s)$ denotes the excess friction kernel
\begin{align}
\gamma_{\mathrm{ex}}(t-s)=\dfrac{1}{k_{\mathrm{B}}T}\dfrac{N\lambda^2\kappa^2_c\kappa^2_b/\gamma^2}{(\kappa_b+\lambda\kappa_c)^2/\gamma^2-4r^2}\dfrac{L^2}{4}\left(e^{-\frac{|t-s|}{\tau_r}}-\dfrac{2r\gamma}{\kappa_b+\lambda\kappa_c}e^{-\frac{|t-s|}{\tau_x}}\right).
\label{ex:excess friction kernel}
\end{align}

In the Markov approximation, the friction kernel can be approximated as
\begin{align}
\gamma(t-s)=2\gamma_{\mathrm{eff}}\delta(t-s),
\end{align}
where $\gamma_{\mathrm{eff}}$ denotes the effective friction coefficient:
\begin{align}
\gamma_{\mathrm{eff}}&:=\int^\infty_0dt\gamma(t)\notag\\
&=\dfrac{N\lambda^2\kappa^2_c}{(\kappa_b+\lambda\kappa_c)^2}\gamma.
\end{align}
For the excess friction kernel, we can show that
\begin{align}
\gamma_{\mathrm{ex}}(t-s)=2\gamma_{\mathrm{ex}}\delta(t-s)
\end{align}
with
\begin{align}
\gamma_{\mathrm{ex}}&:=\int^\infty_0dt\gamma_{\mathrm{ex}}(t)\notag\\
&=\dfrac{1}{k_{\mathrm{B}}T}\dfrac{N\lambda^2\kappa^2_c\kappa^2_bL^2}{8r(\kappa_b+\lambda\kappa_c)^2}.
\end{align}
Thus, in the Markov approximation, the effective generalized Langevin-type equation (\ref{ex:Effective generalized Langevin}) becomes
\begin{align}
M\ddot{X}_t=G(X_t)-(\Gamma+\gamma_{\mathrm{eff}})\dot{X}_t+\sqrt{2(\Gamma+\gamma_{\mathrm{eff}}+\gamma_{\mathrm{ex}})k_{\mathrm{B}}T}\Xi_t.
\label{ex:Effective Langevin}
\end{align}
Note that $\gamma_{\mathrm{ex}}\ge0$.
This implies that stochastic switching enhances the noise intensity.
We remark that if the coupling constant is rescaled as $\lambda=\lambda_0/N^{1/2}$, $\gamma_{\mathrm{eff}}$ and $\gamma_{\mathrm{ex}}$ are finite even in the limit $N\rightarrow\infty$ \cite{maes2014second}.

\subsection{Validity of the effective dynamics}
Because (\ref{ex:Model_particles}) is linear with respect to $x^j_t$, it can be solved exactly.
The stationary solution reads
\begin{align}
x^j_t=\dfrac{\lambda\kappa_c}{\kappa_b+\lambda\kappa_c}X_t+\dfrac{\kappa_b}{\kappa_b+\lambda\kappa_c}\dfrac{L}{2}+\int^t_{-\infty}dse^{-\frac{t-s}{\tau_x}}\left[-\dfrac{\lambda\kappa_c}{\kappa_b+\lambda\kappa_c}\dot{X}_s+\dfrac{\kappa_b}{\gamma}\left(\sigma^j_s-\dfrac{1}{2}\right)L+\sqrt{\dfrac{2k_{\mathrm{B}}T}{\gamma}}\xi^j_s\right]
\label{exact solution}
\end{align}
with $\tau_x=\gamma/(\kappa_b+\lambda\kappa_c)$.
By substituting (\ref{exact solution}) into (\ref{ex:Model_probe}), we obtain
\begin{align}
M\ddot{X}_t&=-\dfrac{N\lambda\kappa_c\kappa_b}{\kappa_b+\lambda\kappa_c}\left(X_t-\dfrac{L}{2}\right)-\Gamma\dot{X}_t-\int^t_{-\infty}dse^{-\frac{t-s}{\tau_x}}\dfrac{N\lambda^2\kappa^2_c}{\kappa_b+\lambda\kappa_c}\dot{X}_s+\sqrt{2\Gamma k_{\mathrm{B}}T}\Xi_t\notag\\
&\quad+\sum_j\lambda\kappa_c\int^t_{-\infty}dse^{-\frac{t-s}{\tau_x}}\left[\dfrac{\kappa_b}{\gamma}\left(\sigma^j_s-\dfrac{1}{2}\right)L+\sqrt{\dfrac{2k_{\mathrm{B}}T}{\gamma}}\xi^j_s\right].
\label{Effective generalized Langevin_exact solution}
\end{align}
Note that the last term corresponds to the noise term $\eta_t=\lambda\kappa_c\sum_j(x^j_t-\langle x^j_t\rangle^{X_t})$ in (\ref{ex:Effective generalized Langevin}) because the exact solution of (\ref{ex:Model_particles}) with $X_t$ held fixed reads
\begin{align}
x^j_t&=\dfrac{\lambda\kappa_c}{\kappa_b+\lambda\kappa_c}X_t+\dfrac{\kappa_b}{\kappa_b+\lambda\kappa_c}\dfrac{L}{2}+\int^t_{-\infty}dse^{-\frac{t-s}{\tau_x}}\left[\dfrac{\kappa_b}{\gamma}\left(\sigma^j_s-\dfrac{1}{2}\right)L+\sqrt{\dfrac{2k_{\mathrm{B}}T}{\gamma}}\xi^j_s\right]\notag\\
&=\langle x^j_t\rangle^{X_t}+\int^t_{-\infty}dse^{-\frac{t-s}{\tau_x}}\left[\dfrac{\kappa_b}{\gamma}\left(\sigma^j_s-\dfrac{1}{2}\right)L+\sqrt{\dfrac{2k_{\mathrm{B}}T}{\gamma}}\xi^j_s\right],
\label{exact solution_X_fixed}
\end{align}
where $\langle x^j_t\rangle^{X_t}=(\lambda\kappa_cX_t+\kappa_bL/2)/(\kappa_b+\lambda\kappa_c)$.
Therefore, (\ref{Effective generalized Langevin_exact solution}) exactly corresponds to (\ref{ex:Effective generalized Langevin}).

We remark that (\ref{ex:Effective Langevin}) can also be obtained by using the singular perturbation method described in Sect.\ \ref{Validity of the effective dynamics}.

\subsection{Fast switching limit}
In the fast switching limit $\tau_r/\tau_x\rightarrow0$ ($r\rightarrow\infty$), we can easily see that the excess friction kernel (\ref{ex:excess friction kernel}) goes to zero:
\begin{align}
\dfrac{1}{k_{\mathrm{B}}T}\langle\eta_t\eta_s\rangle^{X_t}-\gamma(t-s)&=\gamma_{\mathrm{ex}}(t-s)\notag\\
&\rightarrow0,
\end{align}
and thus the standard second FDR is recovered.
Correspondingly, we can show that the upper bound of the inequality for the violation of the second FDR (\ref{bound on the violation of the 2nd FDR}) also goes to zero, as derived below:
\begin{align}
\left|\gamma(t-s)-\dfrac{1}{k_{\mathrm{B}}T}\langle\eta_t\eta_s\rangle^{X_t}\right|&\le\sqrt{\langle\Phi^2\rangle^{X_t}}\sqrt{\mathrm{Var}\left[\partial_{X_t}\Delta s^{X_t}_{\mathrm{tot}}\right]}\notag\\
&\rightarrow0.
\end{align}
The important point here is that, even in this limit, the particles are out of equilibrium.
In fact, if we denote by $\dot{s}^{X_t}_{\mathrm{env}}$ the stochastic entropy production rate of the equilibrium thermal bath for the dynamics (\ref{ex:Model_particles}) with $X_t$ held fixed, it can be shown that
\begin{align}
\langle \dot{s}^{X_t}_{\mathrm{env}}\rangle^{X_t}&=\dfrac{1}{k_{\mathrm{B}}T}\sum_j\langle\dot{x}^j_t\circ[-\kappa_b(x^j_t-\sigma^j_tL)]\rangle^{X_t}\notag\\
&=\dfrac{N}{\gamma k_{\mathrm{B}}T}\dfrac{\kappa^2_bL^2}{2(\kappa_b+\lambda\kappa_c)/r\gamma+4}\notag\\
&\rightarrow\dfrac{N}{k_{\mathrm{B}}T}\dfrac{\kappa^2_bL^2}{4\gamma}\quad\text{as}\quad \tau_r/\tau_x\rightarrow0.
\end{align}
Thus, the potential switching medium (\ref{ex:Model_particles}) in the fast switching limit provides an example of a nonequilibrium medium where the standard second FDR holds.

Note that the entropy production $\langle\dot{s}^{X_t}_{\mathrm{env}}\rangle^{X_t}$ is induced by the fast degrees of freedom $\sigma^j$, which does not appear in the dynamics in the fast switching limit:
\begin{align}
\gamma\dot{x}^j_t=-\kappa_b\left(x^j_t-\dfrac{L}{2}\right)-\lambda\kappa_c(x^j_t-X_t)+\sqrt{2\gamma k_{\mathrm{B}}T}\xi^j_t.
\label{ex:Model_particles_fast_switching_limit}
\end{align}
(\ref{ex:Model_particles_fast_switching_limit}) can, for example, be obtained by using the singular perturbation method as described in Sect.\ \ref{Validity of the effective dynamics}.
Therefore, the potential switching medium appears to be just an equilibrium thermal bath, and thus the standard second FDR holds.
We remark that $\langle\dot{s}^{X_t}_{\mathrm{env}}\rangle^{X_t}$ is an example of \textit{hidden entropy}, i.e., an entropy production invisible from the coarse-grained dynamics \cite{celani2012anomalous,kawaguchi2013fluctuation,wang2016entropy}.

We now show that the upper bound of the inequality (\ref{bound on the violation of the 2nd FDR}) goes to zero in the fast switching limit.
To this end, we calculate $\langle(\Phi({\bm x}_t,X_t))^2\rangle^{X_t}$ and $\mathrm{Var}\left[\partial_{X_t}\Delta s^{X_t}_{\mathrm{tot}}\right]$.
By using (\ref{exact solution_X_fixed}) and (\ref{appendix: xx_correlation}), $\langle(\Phi({\bm x}_t,X_t))^2\rangle^{X_t}$ can be calculated as
\begin{align}
\langle(\Phi({\bm x}_t,X_t))^2\rangle^{X_t}&=\lambda^2\kappa^2_c\sum_i\sum_j\langle(x^i_t-X_t)(x^j_t-X_t)\rangle^{X_t}\notag\\
&=\dfrac{N^2\lambda^2\kappa^2_c\kappa^2_b}{(\kappa_b+\lambda\kappa_c)^2}\left(X_t-\dfrac{L}{2}\right)^2+N\lambda^2\kappa^2_c\left[\dfrac{k_{\mathrm{B}}T}{\kappa_b+\lambda\kappa_c}+\dfrac{\kappa^2_b/\gamma^2}{(\kappa_b+\lambda\kappa_c)^2/\gamma^2-4r^2}\dfrac{L^2}{4}\left(1-\dfrac{2r\gamma}{\kappa_b+\lambda\kappa_c}\right)\right].
\end{align}
From this expression, it follows that
\begin{align}
\langle(\Phi({\bm x}_t,X_t))^2\rangle^{X_t}\rightarrow\dfrac{N^2\lambda^2\kappa^2_c\kappa^2_b}{(\kappa_b+\lambda\kappa_c)^2}\left(X_t-\dfrac{L}{2}\right)^2+Nk_{\mathrm{B}}T\dfrac{\lambda^2\kappa^2_c}{\kappa_b+\lambda\kappa_c}
\label{upper_bound_phi_fast_switching_limit}
\end{align}
in the fast switching limit.
$\mathrm{Var}\left[\partial_{X_t}\Delta s^{X_t}_{\mathrm{tot}}\right]$ can be explicitly calculated from the expression (\ref{entropy_reponse}) as follows:
\begin{align}
\mathrm{Var}\left[\partial_{X_t}\Delta s^{X_t}_{\mathrm{tot}}\right]&=2\mathrm{Var}\left[-\partial_{X_t}\ln P^{X_t}_{\mathrm{ss}}({\bm x}_s,{\bm\sigma}_s)+\dfrac{1}{k_{\mathrm{B}}T}\Phi({\bm x}_s,X_t)\right]\notag\\
&=2N\left(\dfrac{\lambda\kappa_c}{k_{\mathrm{B}}T}\right)^2\left[\dfrac{k_{\mathrm{B}}T}{\kappa_b+\lambda\kappa_c}+\dfrac{\kappa^2_b/\gamma^2}{(\kappa_b+\lambda\kappa_c)^2/\gamma^2-4r^2}\dfrac{L^2}{4}\left(1-\dfrac{2r\gamma}{\kappa_b+\lambda\kappa_c}\right)\right]\notag\\
&\quad-4N\left(\dfrac{\lambda\kappa_c}{k_{\mathrm{B}}T}\right)^2\dfrac{k_{\mathrm{B}}T}{\kappa_b+\lambda\kappa_c}+2I(X_t),
\label{upper_bound_S}
\end{align}
where $I(X_t)$ denotes the Fisher information \cite{cover1999elements} defined by
\begin{equation}
I(X_t):=\langle(\partial_{X_t}\ln P^{X_t}_{\mathrm{ss}}({\bm x}_t,{\bm \sigma}_t))^2\rangle^{X_t}.
\end{equation}
By noting that, in the fast switching limit, $P^{X_t}_{\mathrm{ss}}({\bm x},{\bm \sigma})$ is given by
\begin{equation}
P^{X_t}_{\mathrm{ss}}({\bm x},{\bm \sigma})=\mathcal{N}\exp\left(-\dfrac{1}{k_{\mathrm{B}}T}\sum_j\left[\dfrac{\kappa_b}{2}\left(x^j-\dfrac{L}{2}\right)^2+\dfrac{\lambda\kappa_c}{2}\left(x^j-X_t\right)^2\right]\right),
\end{equation}
where
\begin{align}
\mathcal{N}=\dfrac{1}{2^N}\left(\dfrac{\kappa_b+\lambda\kappa_c}{2\pi k_{\mathrm{B}}T}\right)^{N/2}\exp\left(\dfrac{N}{2k_{\mathrm{B}}T}\dfrac{\lambda\kappa_c\kappa_b}{\kappa_b+\lambda\kappa_c}\left(X_t-\dfrac{L}{2}\right)^2\right),
\end{align}
we obtain
\begin{align}
I(X_t)&=N\left(\dfrac{\lambda\kappa_c}{k_{\mathrm{B}}T}\right)^2\left[\dfrac{k_{\mathrm{B}}T}{\kappa_b+\lambda\kappa_c}+\dfrac{\kappa^2_b/\gamma^2}{(\kappa_b+\lambda\kappa_c)^2/\gamma^2-4r^2}\dfrac{L^2}{4}\left(1-\dfrac{2r\gamma}{\kappa_b+\lambda\kappa_c}\right)\right].
\label{I_fast_switching_limit}
\end{align} 
By substituting (\ref{I_fast_switching_limit}) into (\ref{upper_bound_S}), we find that in the limit $\tau_r/\tau_x\rightarrow0$,
\begin{align}
\mathrm{Var}\left[\partial_{X_t}\Delta s^{X_t}_{\mathrm{tot}}\right]&=4N\left(\dfrac{\lambda\kappa_c}{k_{\mathrm{B}}T}\right)^2\left[\dfrac{\kappa^2_b/\gamma^2}{(\kappa_b+\lambda\kappa_c)^2/\gamma^2-4r^2}\dfrac{L^2}{4}\left(1-\dfrac{2r\gamma}{\kappa_b+\lambda\kappa_c}\right)\right]\notag\\
&\rightarrow0.
\label{upper_bound_S_fast_switching_limit}
\end{align}
From (\ref{upper_bound_phi_fast_switching_limit}) and (\ref{upper_bound_S_fast_switching_limit}), we thus find that the upper bound of the inequality (\ref{bound on the violation of the 2nd FDR}) goes to zero in the fast switching limit.

\section{Concluding remarks\label{Concluding remarks}}
In summary, we have investigated a class of nonequilibrium media described by Langevin dynamics that satisfies the LDB.
For the effective dynamics of a probe immersed in the medium, we have derived an inequality that bounds the violation of the second FDR.
The upper bound of the inequality can be interpreted as a measure of robustness of the nonequilibrium medium against perturbation of the probe position.
This implies that a nonequilibrium medium may be characterized by robustness against perturbation.
We have also discussed the validity of the effective dynamics.
In particular, we have shown that the effective dynamics obtained from nonequilibrium linear response theory is consistent with that obtained from the singular perturbation method.
As an example of these results, we have proposed the potential switching medium in which the particles are subjected to potentials that switch stochastically.
For this model, we have shown that the second FDR is recovered in the fast switching limit, although the particles are out of equilibrium.

Although we have focused on a class of nonequilibrium media described by Langevin dynamics that satisfies the LDB, it is possible to derive effective dynamics for more general nonequilibrium media.
For example, Maes has recently derived the effective dynamics of a probe immersed in an active Ornstein-Uhlenbeck (AOU) medium \cite{maes2020fluctuating,martin2021statistical}.
In that case, the persistence of the medium generates extra mass and additional friction breaking the second FDR.
Because this violation of the second FDR also originates from the violation of the first FDR, we expect that a relation similar to (\ref{bound on the violation of the 2nd FDR}) still holds even for this case.
We also remark that the singular perturbation method described in this paper can be applied to more general nonequilibrium media, including the AOU medium.

\begin{acknowledgements}
The author thanks Ken Hiura for helpful discussions on the singular perturbation method.
The author also thanks Shin-ichi Sasa for valuable comments throughout the manuscript.
The present study was supported by JSPS KAKENHI Grant No. 20J20079, a Grant-in-Aid for JSPS Fellows.
\end{acknowledgements}

\appendix\normalsize
\renewcommand{\theequation}{\Alph{section}.\arabic{equation}}
\setcounter{equation}{0}
\makeatletter
  \def\@seccntformat#1{%
    \@nameuse{@seccnt@prefix@#1}%
    \@nameuse{the#1}%
    \@nameuse{@seccnt@postfix@#1}%
    \@nameuse{@seccnt@afterskip@#1}}
  \def\@seccnt@prefix@section{Appendix }
  \def\@seccnt@postfix@section{:}
  \def\@seccnt@afterskip@section{\ }
  \def\@seccnt@afterskip@subsection{\ }
\makeatother

\section{Derivation of the effective dynamics for the potential switching medium\label{Explicit calculation}}
\subsection{Excess action}
We first confirm that, to first order in $X_s-X_t$, the excess action $\mathcal{A}([{\bm x},{\bm\sigma}]|[X]))$ is given by
\begin{align}
-\mathcal{A}([{\bm x},{\bm \sigma}]|[X])\simeq\dfrac{1}{2k_{\mathrm{B}}T}\left[\int^t_{-\infty}ds(X_s-X_t)\sum_j\dfrac{\partial}{\partial x^j_s}\Phi({\bm x}_s,X_t)\circ\dot{x}^j_s-\int^t_{-\infty}ds(X_s-X_t)\mathcal{L}^\dagger_s\Phi({\bm x}_s,X_t)\right]
\end{align}
with the backward operator for the dynamics (\ref{ex:Model_particles}) with $X_t$ held fixed:
\begin{align}
\mathcal{L}^\dagger_s:=\sum_j\left[\dfrac{1}{\gamma}(-\kappa_b(x^j_s-\sigma^j_sL)-\lambda\kappa_c(x^j_s-X_t))\dfrac{\partial}{\partial x^j_s}+\dfrac{k_{\mathrm{B}}T}{\gamma}\dfrac{\partial^2}{\partial (x^j_s)^2}\right].
\end{align}
To this end, we calculate $\mathbb{P}([{\bm x},{\bm \sigma}]|[X])$ and $\mathbb{P}([{\bm x},{\bm \sigma}]|X_t)$.
We first consider a trajectory in the time interval $[0,t]$ and discretized time $t_n=n\Delta t\in[0,t]$ ($n=0,1,\cdots,M$) with $t\equiv M\Delta t$.
Correspondingly, let $[{\bm x},{\bm \sigma}]:=\{({\bm x}_0,{\bm \sigma}_0),({\bm x}_1,{\bm \sigma}_1),\cdots,({\bm x}_M,{\bm \sigma}_M)\}$ be the discretized trajectory, where $({\bm x}_n,{\bm\sigma}_n):=({\bm x}_{t_n},{\bm\sigma}_{t_n})$.
Suppose that the state $\sigma^j$ is switched at time intervals with $n=n^j_1,n^j_2,\cdots,n^j_{k_j}\in\{0,1,\cdots,M\}$ as
\begin{align}
\sigma^j_{n^j_\ell+1}=1-\sigma^j_{n^j_\ell}.
\end{align}
We denote by $\Sigma^j_\ell\in\{0,1\}$ the value of $\sigma^j_n$ for $n^j_\ell<n\le n^j_{\ell+1}$ with $n^j_0:=-1$ and $n^j_{k_j+1}:=M$.
For notational simplicity, we rewrite (\ref{ex:Model_particles}) as
\begin{align}
\gamma\dot{x}^j_s&=-U'_1(x^j_s,\sigma^j_s)-\lambda V'_1(x^j_s,X_s)+\sqrt{2\gamma k_{\mathrm{B}}T}\xi^j_s\notag\\
&=-U'_1(x^j_s,\sigma^j_s)-\lambda V'_1(x^j_s,X_t)+h_sf(x^j_s)+\sqrt{2\gamma k_{\mathrm{B}}T}\xi^j_s,
\end{align}
where $U_1(x^j,\sigma^j):=\kappa_b(x^j-\sigma^jL)^2/2$, $V_1(x^j,X):=\kappa_c(x^j-X)^2/2$, and the prime denotes the derivative with respect to $x^j_t$.
In the second line, we have introduced a time-dependent amplitude $h_s:=X_s-X_t$ and $f(x^j_s):=\partial_j\Phi({\bm x}_s,X_t)=\lambda\kappa_c$ to explicitly represent the deviation from the dynamics with $X_t$ held fixed.
Then, the probability density of a trajectory $\mathbb{P}([{\bm x},{\bm \sigma}]|[X])$ starting from $({\bm x}_0,{\bm \sigma}_0)$ reads \cite{harada2007fluctuations}
\begin{align}
&\quad\mathbb{P}([{\bm x},{\bm \sigma}]|[X])\notag\\
&=\prod_j\prod^{n^j_1-1}_{n=0}\sqrt{\dfrac{\gamma}{4\pi k_{\mathrm{B}}T\Delta t}}e^{-\frac{\Delta t}{4\gamma k_{\mathrm{B}}T}\left[\gamma\frac{x^j_{n+1}-x^j_n}{\Delta t}+U'_1(\bar{x}^j_n,\Sigma^j_0)+\lambda V'_1(\bar{x}^j_n,X_t)-\bar{h}_nf(\bar{x}^j_n)\right]^2+\frac{\Delta t}{2\gamma}\left[U''_1(\bar{x}^j_n,\Sigma^j_0)+\lambda V''_1(\bar{x}^j_n,X_t)-\bar{h}_nf'(\bar{x}^j_n)\right]-r\Delta t}\notag\\
&\times\prod^{k_j}_{\ell=1}r\Delta t\delta(x^j_{n^j_\ell+1}-x^j_{n^j_\ell})\notag\\
&\times\prod^{n^j_{\ell+1}-1}_{n=n^j_\ell+1}\sqrt{\dfrac{\gamma}{4\pi k_{\mathrm{B}}T\Delta t}}e^{-\frac{\Delta t}{4\gamma k_{\mathrm{B}}T}\left[\gamma\frac{x^j_{n+1}-x^j_n}{\Delta t}+U'_1(\bar{x}^j_n,\Sigma^j_\ell)+\lambda V'_1(\bar{x}^j_n,X_t)-\bar{h}_nf(\bar{x}^j_n)\right]^2+\frac{\Delta t}{2\gamma}\left[U''_1(\bar{x}^j_n,\Sigma^j_\ell)+\lambda V''_1(\bar{x}^j_n,X_t)-\bar{h}_nf'(\bar{x}^j_n)\right]-r\Delta t}.
\label{path probability}
\end{align}
Here, $\bar{x}^j_n:=(x^j_{n+1}+x^j_n)/2$ and $\bar{h}_n:=(h_{n+1}+h_n)/2$.
We note that $\mathbb{P}([{\bm x},{\bm \sigma}]|X_t)$ is immediately obtained from (\ref{path probability}) by setting $\bar{h}_n=0$.
From these expressions, it follows that
\begin{align}
&\quad\ln\dfrac{\mathcal{\mathbb{P}}([{\bm x},{\bm \sigma}]|[X])}{\mathbb{P}([{\bm x},{\bm \sigma}]|X_t)}\notag\\
&=\sum_j\left[\dfrac{1}{2k_{\mathrm{B}}T}\sum^{M-1}_{n=0}\bar{h}_nf(\bar{x}^j_n)(x^j_{n+1}-x^j_n)-\dfrac{1}{2\gamma k_{\mathrm{B}}T}\left\{\sum^{n^j_1-1}_{n=0}\bar{h}_nf(\bar{x}^j_n)[-U'_1(\bar{x}^j_n,\Sigma^j_0)-\lambda V'_1(\bar{x}^j_n,X_t)]\Delta t\right.\right.\notag\\
&\quad\left.\left.+\sum^{k_j}_{\ell=1}\sum^{n^j_{\ell+1}-1}_{n=n^j_\ell+1}\bar{h}_nf(\bar{x}^j_n)[-U'_1(\bar{x}^j_n,\Sigma^j_\ell)-\lambda V'_1(\bar{x}^j_n,X_t)]\Delta t\right\}-\dfrac{1}{2\gamma}\sum^{M-1}_{n=0}\bar{h}_nf'(\bar{x}^j_n)\Delta t\right]+O(\bar{h}^2_n).
\end{align}
By taking the continuum limit and replacing the time interval from $[0,t]$ to $[-\infty,t]$, we obtain the excess action $\mathcal{A}([{\bm x},{\bm \sigma}]|[X])$:
\begin{align}
-\mathcal{A}([{\bm x},{\bm \sigma}]|[X])&=\ln\dfrac{\mathbb{P}([{\bm x},{\bm \sigma}]|[X])}{\mathbb{P}([{\bm x},{\bm \sigma}]|X_t)}\notag\\
&=\sum_j\left[\dfrac{1}{2k_{\mathrm{B}}T}\int^t_{-\infty}dsh_sf(x^j_s)\circ\dot{x}^j_s\right.\notag\\
&\left.-\dfrac{1}{2\gamma k_{\mathrm{B}}T}\int^t_{-\infty}dsh_sf(x^j_s)(-U'_1(x^j_s,\sigma^j_s)-\lambda V'_1(x^j_s,X_t))-\dfrac{1}{2\gamma}\int^t_{-\infty}dsh_s\dfrac{\partial}{\partial x^j_s}f(x^j_s)\right]+O(h^2_s)\notag\\
&\simeq\dfrac{1}{2k_{\mathrm{B}}T}\left[\int^t_{-\infty}dsh_s\sum_j\dfrac{\partial}{\partial x^j_s}\Phi({\bm x}_s,X_t)\circ\dot{x}^j_s-\int^t_{-\infty}dsh_s\mathcal{L}^\dagger_s\Phi({\bm x}_s,X_t)\right],
\end{align}
where
\begin{align}
\mathcal{L}^\dagger_s:=\sum_j\left[\dfrac{1}{\gamma}(-U'_1(x^j_s,\sigma^j_s)-\lambda V'_1(x^j_s,X_t))\dfrac{\partial}{\partial x^j_s}+\dfrac{k_{\mathrm{B}}T}{\gamma}\dfrac{\partial^2}{\partial (x^j_s)^2}\right].
\end{align}

\subsection{Explicit calculation of $G(X_t)$, $\gamma(t-s)$, and $\gamma_{\mathrm{ex}}(t-s)$}
Here, we calculate $G(X_t)$, $\gamma(t-s)$, and $\gamma_{\mathrm{ex}}(t-s)$ explicitly.
The starting point is the stationary solution of (\ref{ex:Model_particles}) with $X_t$ held fixed (\ref{exact solution_X_fixed}):
\begin{align}
x^j_t&=\dfrac{\lambda\kappa_c}{\kappa_b+\lambda\kappa_c}X_t+\dfrac{\kappa_b}{\kappa_b+\lambda\kappa_c}\dfrac{L}{2}+\int^t_{-\infty}dse^{-\frac{t-s}{\tau_x}}\left[\dfrac{\kappa_b}{\gamma}\left(\sigma^j_s-\dfrac{1}{2}\right)L+\sqrt{\dfrac{2k_{\mathrm{B}}T}{\gamma}}\xi^j_s\right]\notag\\
&=\langle x^j_t\rangle^{X_t}+\int^t_{-\infty}dse^{-\frac{t-s}{\tau_x}}\left[\dfrac{\kappa_b}{\gamma}\left(\sigma^j_s-\dfrac{1}{2}\right)L+\sqrt{\dfrac{2k_{\mathrm{B}}T}{\gamma}}\xi^j_s\right].
\label{appendix: exact solution_X_fixed}
\end{align}

The statistical force is immediately obtained by substituting (\ref{appendix: exact solution_X_fixed}) into its definition:
\begin{align}
G(X_t)&:=\langle\Phi({\bm x}_t,X_t)\rangle^{X_t}\notag\\
&=\left\langle-\lambda\kappa_c\sum_j(X_t-x^j_t)\right\rangle^{X_t}\notag\\
&=-\dfrac{N\lambda\kappa_c\kappa_b}{\kappa_b+\lambda\kappa_c}\left(X_t-\dfrac{L}{2}\right).
\end{align}

To calculate the friction kernel, we first calculate the response function $R_{\Phi\Phi}(t-s)$.
The response function $R_{\Phi\Phi}(t-s)$ is expressed as
\begin{align}
&R_{\Phi\Phi}(t-s)\notag\\
&=\dfrac{1}{2k_{\mathrm{B}}T}\left[\dfrac{d}{ds}\langle \Phi({\bm x}_s,X_t);\Phi({\bm x}_t,X_t)\rangle^{X_t}-\langle\mathcal{L}^\dagger_s\Phi({\bm x}_s,X_t);\Phi({\bm x}_t,X_t)\rangle^{X_t}\right]\notag\\
&=\dfrac{\lambda^2\kappa^2_c}{2k_{\mathrm{B}}T}\sum_i\sum_j\left[\dfrac{d}{ds}\langle x^i_s-X_t;x^j_t-X_t\rangle^{X_t}+\dfrac{\kappa_b}{\gamma}\langle x^i_s-\sigma^i_sL;x^j_t-X_t\rangle^{X_t}+\dfrac{\lambda\kappa_c}{\gamma}\langle x^i_s-X_t;x^j_t-X_t\rangle^{X_t}\right]\notag\\
&=\dfrac{\lambda^2\kappa^2_c}{2k_{\mathrm{B}}T}\sum_i\sum_j\left[\dfrac{d}{ds}\langle x^i_s;x^j_t\rangle^{X_t}+\dfrac{\kappa_b+\lambda\kappa_c}{\gamma}\langle x^i_s;x^j_t\rangle^{X_t}-\dfrac{\kappa_b}{\gamma}L\langle \sigma^i_s;x^j_t\rangle^{X_t}\right].
\end{align}
By using (\ref{appendix: exact solution_X_fixed}) and the relation
\begin{align}
\langle\sigma^i_t\sigma^j_s\rangle=
\begin{cases}
(1+e^{-2r|t-s|})/4\quad\text{for}\quad i=j\notag\\
1/4\quad\text{for}\quad i\neq j,
\end{cases}
\end{align}
$\langle x^i_s;x^j_t\rangle^{X_t}$ and $\langle \sigma^i_s;x^j_t\rangle^{X_t}$ are calculated as
\begin{align}
\langle x^i_s;x^j_t\rangle^{X_t}
&=\delta_{ij}\left[\dfrac{\kappa^2_b/\gamma^2}{(\kappa_b+\lambda\kappa_c)^2/\gamma^2-4r^2}\dfrac{L^2}{4}\left(e^{-2r|t-s|}-\dfrac{2r\gamma}{\kappa_b+\lambda\kappa_c}e^{-\frac{\kappa_b+\lambda\kappa_c}{\gamma}|t-s|}\right)+\dfrac{k_{\mathrm{B}}T}{\kappa_b+\lambda\kappa_c}e^{-\frac{\kappa_b+\lambda\kappa_c}{\gamma}|t-s|}\right],
\label{appendix: xx_correlation}
\end{align}
\begin{align}
&\langle \sigma^i_s;x^j_t\rangle^{X_t}\notag\\
&=\delta_{ij}
\begin{cases}
\dfrac{\kappa_b/\gamma}{(\kappa_b+\lambda\kappa_c)/\gamma+2r}\dfrac{L}{4}e^{-2r|t-s|},\quad\text{for}\quad t<s\\
\dfrac{\kappa_b/\gamma}{(\kappa_b+\lambda\kappa_c)/\gamma-2r}\dfrac{L}{4}\left(e^{-2r|t-s|}-e^{-\frac{\kappa_b+\lambda\kappa_c}{\gamma}|t-s|}\right)+\dfrac{\kappa_b/\gamma}{(\kappa_b+\lambda\kappa_c)/\gamma+2r}\dfrac{L}{4}e^{-\frac{\kappa_b+\lambda\kappa_c}{\gamma}|t-s|},\quad\text{for}\quad t\ge s.
\end{cases}
\end{align}
Therefore, for $t\ge s$, the response function is
\begin{align}
R_{\Phi\Phi}(t-s)=\dfrac{N\lambda^2\kappa^2_c}{\gamma}e^{-\frac{\kappa_b+\lambda\kappa_c}{\gamma}(t-s)}.
\end{align}
From this result, it follows that
\begin{align}
\gamma(t-s)&:=\int^s_{-\infty}du R_{\Phi\Phi}(t-u)\notag\\
&=\dfrac{N\lambda^2\kappa^2_c}{\kappa_b+\lambda\kappa_c}e^{-\frac{\kappa_b+\lambda\kappa_c}{\gamma}(t-s)},\quad\text{for}\quad t\ge s.
\end{align}

To obtain the explicit expression of $\gamma_{\mathrm{ex}}(t-s)$, we calculate the noise correlation $\langle\eta_t\eta_s\rangle^{X_t}$.
By using (\ref{appendix: xx_correlation}), the noise correlation is calculated as
\begin{align}
\langle\eta_t\eta_s\rangle^{X_t}&=\lambda^2\kappa^2_c\sum_i\sum_j\langle(x^i_t-X_t);(x^j_s-X_t)\rangle^{X_t}\notag\\
&=k_{\mathrm{B}}T\gamma(t-s)+\dfrac{N\lambda^2\kappa^2_c\kappa^2_b/\gamma^2}{(\kappa_b+\lambda\kappa_c)^2/\gamma^2-4r^2}\dfrac{L^2}{4}\left(e^{-2r|t-s|}-\dfrac{2r\gamma}{\kappa_b+\lambda\kappa_c}e^{-\frac{\kappa_b+\lambda\kappa_c}{\gamma}|t-s|}\right).
\end{align}
Thus, the excess friction kernel $\gamma_{\mathrm{ex}}(t-s)$ is given by
\begin{align}
\gamma_{\mathrm{ex}}(t-s)=\dfrac{1}{k_{\mathrm{B}}T}\dfrac{N\lambda^2\kappa^2_c\kappa^2_b/\gamma^2}{(\kappa_b+\lambda\kappa_c)^2/\gamma^2-4r^2}\dfrac{L^2}{4}\left(e^{-2r|t-s|}-\dfrac{2r\gamma}{\kappa_b+\lambda\kappa_c}e^{-\frac{\kappa_b+\lambda\kappa_c}{\gamma}|t-s|}\right).
\end{align}

%
%

\bibliographystyle{spmpsci_unsort}      
\bibliography{2nd_FDT}   

\begin{thebibliography}{10}
\providecommand{\url}[1]{{#1}}
\providecommand{\urlprefix}{URL }
\expandafter\ifx\csname urlstyle\endcsname\relax
  \providecommand{\doi}[1]{DOI~\discretionary{}{}{}#1}\else
  \providecommand{\doi}{DOI~\discretionary{}{}{}\begingroup
  \urlstyle{rm}\Url}\fi

\bibitem{kubo1966fluctuation}
Kubo, R.: The fluctuation-dissipation theorem.
\newblock Rep. Prog. Phys. \textbf{29}(1), 255 (1966)

\bibitem{kubo2012statistical}
Kubo, R., Toda, M., Hashitsume, N.: Statistical physics {II}: nonequilibrium
  statistical mechanics, vol.~31.
\newblock Springer-Verlag, Berlin (2012)

\bibitem{harada2005phenomenological}
Harada, T.: Phenomenological energetics for molecular motors.
\newblock Europhys. Lett. \textbf{70}(1), 49 (2005)

\bibitem{harada2005equality}
Harada, T., Sasa, S.i.: Equality {C}onnecting {E}nergy {D}issipation with a
  {V}iolation of the {F}luctuation-{R}esponse {R}elation.
\newblock Phys. Rev. Lett. \textbf{95}(13), 130602 (2005)

\bibitem{harada2006energy}
Harada, T., Sasa, S.i.: Energy dissipation and violation of the
  fluctuation-response relation in nonequilibrium {L}angevin systems.
\newblock Phys. Rev. E \textbf{73}(2), 026131 (2006)

\bibitem{harada2007fluctuations}
Harada, T., Sasa, S.i.: Fluctuations, responses and energetics of molecular
  motors.
\newblock Math. Biosci. \textbf{207}(2), 365--386 (2007)

\bibitem{cugliandolo1997fluctuation}
Cugliandolo, L.F., Dean, D.S., Kurchan, J.: Fluctuation-{D}issipation
  {T}heorems and {E}ntropy {P}roduction in {R}elaxational {S}ystems.
\newblock Phys. Rev. Lett. \textbf{79}(12), 2168 (1997)

\bibitem{toyabe2015single}
Toyabe, S., Muneyuki, E.: Single molecule thermodynamics of {ATP} synthesis by
  {F1-ATP}ase.
\newblock New J. Phys. \textbf{17}(1), 015008 (2015)

\bibitem{ariga2018nonequilibrium}
Ariga, T., Tomishige, M., Mizuno, D.: Nonequilibrium energetics of molecular
  motor kinesin.
\newblock Phys. Rev. Lett. \textbf{121}(21), 218101 (2018)

\bibitem{matsumoto2014response}
Matsumoto, T., Otsuki, M., Ooshida, T., Goto, S., Nakahara, A.: Response
  function of turbulence computed via fluctuation-response relation of a
  {L}angevin system with vanishing noise.
\newblock Phys. Rev. E \textbf{89}(6), 061002 (2014)

\bibitem{matsumoto2021correlation}
Matsumoto, T., Otsuki, M., Ooshida, T., Goto, S.: Correlation function and
  linear response function of homogeneous isotropic turbulence in the
  {E}ulerian and {L}agrangian coordinates.
\newblock J. Fluid Mech. \textbf{919} (2021)

\bibitem{maes2014second}
Maes, C.: On the {S}econd {F}luctuation--{D}issipation {T}heorem for
  {N}onequilibrium {B}aths.
\newblock J. Stat. Phys. \textbf{154}(3), 705--722 (2014)

\bibitem{hayashi2006linear}
Hayashi, K., Sasa, S.i.: Linear response theory in stochastic many-body systems
  revisited.
\newblock Phys. A \textbf{370}(2), 407--429 (2006)

\bibitem{maes2013fluctuation}
Maes, C., Safaverdi, S., Visco, P., Van~Wijland, F.: Fluctuation-response
  relations for nonequilibrium diffusions with memory.
\newblock Phys. Rev. E \textbf{87}(2), 022125 (2013)

\bibitem{maes2020response}
Maes, C.: Response theory: a trajectory-based approach.
\newblock Front. Phys. \textbf{8}, 229 (2020)

\bibitem{maes2021local}
Maes, C.: Local detailed balance.
\newblock SciPost Phys. p.~32 (2021)

\bibitem{mizuno2007nonequilibrium}
Mizuno, D., Tardin, C., Schmidt, C.F., MacKintosh, F.C.: Nonequilibrium
  mechanics of active cytoskeletal networks.
\newblock Science \textbf{315}(5810), 370--373 (2007)

\bibitem{nishizawa2017feedback}
Nishizawa, K., Bremerich, M., Ayade, H., Schmidt, C.F., Ariga, T., Mizuno, D.:
  Feedback-tracking microrheology in living cells.
\newblock Sci. Adv. \textbf{3}(9), e1700318 (2017)

\bibitem{jee2018catalytic}
Jee, A.Y., Cho, Y.K., Granick, S., Tlusty, T.: Catalytic enzymes are active
  matter.
\newblock Proc. Natl. Acad. Sci. \textbf{115}(46), E10812--E10821 (2018)

\bibitem{ross2016dark}
Ross, J.L.: The dark matter of biology.
\newblock Biophys. J. \textbf{111}(5), 909--916 (2016)

\bibitem{ariga2021noise}
Ariga, T., Tateishi, K., Tomishige, M., Mizuno, D.: Noise-induced acceleration
  of single molecule kinesin-1.
\newblock Phys. Rev. Lett. \textbf{127}(17), 178101 (2021)

\bibitem{hayashi2006law}
Hayashi, K., Sasa, S.i.: The law of action and reaction for the effective force
  in a non-equilibrium colloidal system.
\newblock J. Phys. \textbf{18}(10), 2825 (2006)

\bibitem{maes2015friction}
Maes, C., Steffenoni, S.: Friction and noise for a probe in a nonequilibrium
  fluid.
\newblock Phys. Rev. E \textbf{91}(2), 022128 (2015)

\bibitem{haga2015nonequilibrium}
Haga, T.: Nonequilibrium {L}angevin equation and effective temperature for
  particle interacting with spatially extended environment.
\newblock J. Stat. Phys. \textbf{159}(3), 713--729 (2015)

\bibitem{steffenoni2016interacting}
Steffenoni, S., Kroy, K., Falasco, G.: Interacting {B}rownian dynamics in a
  nonequilibrium particle bath.
\newblock Phys. Rev. E \textbf{94}(6), 062139 (2016)

\bibitem{maes2017induced}
Maes, C., Thiery, T.: The induced motion of a probe coupled to a bath with
  random resettings.
\newblock J. Phys. A \textbf{50}(41), 415001 (2017)

\bibitem{baiesi2009fluctuations}
Baiesi, M., Maes, C., Wynants, B.: Fluctuations and {R}esponse of
  {N}onequilibrium {S}tates.
\newblock Phys. Rev. Lett. \textbf{103}(1), 010602 (2009)

\bibitem{baiesi2009nonequilibrium}
Baiesi, M., Maes, C., Wynants, B.: Nonequilibrium linear response for markov
  dynamics, {I}: jump processes and overdamped diffusions.
\newblock J. Stat. Phys. \textbf{137}(5), 1094--1116 (2009)

\bibitem{baiesi2013update}
Baiesi, M., Maes, C.: An update on the nonequilibrium linear response.
\newblock New J. Phys. \textbf{15}(1), 013004 (2013)

\bibitem{maes2020fluctuating}
Maes, C.: Fluctuating {M}otion in an {A}ctive {E}nvironment.
\newblock Phys. Rev. Lett. \textbf{125}(20), 208001 (2020)

\bibitem{toyabe2007experimental}
Toyabe, S., Jiang, H.R., Nakamura, T., Murayama, Y., Sano, M.: Experimental
  test of a new equality: {M}easuring heat dissipation in an optically driven
  colloidal system.
\newblock Phys. Rev. E \textbf{75}(1), 011122 (2007)

\bibitem{dieterich2015single}
Dieterich, E., Camunas-Soler, J., Ribezzi-Crivellari, M., Seifert, U., Ritort,
  F.: Single-molecule measurement of the effective temperature in
  non-equilibrium steady states.
\newblock Nat. Phys. \textbf{11}(11), 971--977 (2015)

\bibitem{wang2016entropy}
Wang, S.W., Kawaguchi, K., Sasa, S.i., Tang, L.H.: Entropy production of
  nanosystems with time scale separation.
\newblock Phys. Rev. Lett. \textbf{117}(7), 070601 (2016)

\bibitem{celani2012anomalous}
Celani, A., Bo, S., Eichhorn, R., Aurell, E.: Anomalous thermodynamics at the
  microscale.
\newblock Phys. Rev. Lett. \textbf{109}(26), 260603 (2012)

\bibitem{kawaguchi2013fluctuation}
Kawaguchi, K., Nakayama, Y.: Fluctuation theorem for hidden entropy production.
\newblock Phys. Rev. E \textbf{88}(2), 022147 (2013)

\bibitem{zwanzig1973nonlinear}
Zwanzig, R.: Nonlinear generalized {L}angevin equations.
\newblock J. Stat. Phys. \textbf{9}(3), 215--220 (1973)

\bibitem{zwanzig2001nonequilibrium}
Zwanzig, R.: Nonequilibrium {S}tatistical {M}echanics.
\newblock Oxford University Press (2001)

\bibitem{girsanov1960transforming}
Girsanov, I.V.: On transforming a certain class of stochastic processes by
  absolutely continuous substitution of measures.
\newblock Theory Probab. Appl. \textbf{5}(3), 285--301 (1960)

\bibitem{gardiner1985handbook}
Gardiner, C.W.: Handbook of {S}tochastic {M}ethods, 4th edn.
\newblock Springer, Berlin (2009)

\bibitem{maes2020frenesy}
Maes, C.: Frenesy: {T}ime-symmetric dynamical activity in nonequilibria.
\newblock Phys. Rep. \textbf{850}, 1--33 (2020)

\bibitem{risken1996fokker}
Risken, H.: The {F}okker-{P}lanck {E}quation.
\newblock Springer (1996)

\bibitem{seifert2010fluctuation}
Seifert, U., Speck, T.: Fluctuation-dissipation theorem in nonequilibrium
  steady states.
\newblock Europhys. Lett. \textbf{89}(1), 10007 (2010)

\bibitem{seifert2005entropy}
Seifert, U.: Entropy production along a stochastic trajectory and an integral
  fluctuation theorem.
\newblock Phys. Rev. Lett. \textbf{95}(4), 040602 (2005)

\bibitem{sekimoto2010stochastic}
Sekimoto, K.: Stochastic Energetics.
\newblock Springer, New York (2010)

\bibitem{seifert2012stochastic}
Seifert, U.: Stochastic thermodynamics, fluctuation theorems and molecular
  machines.
\newblock Rep. Prog. Phys. \textbf{75}(12), 126001 (2012)

\bibitem{zwanzig1972memory}
Zwanzig, R., Nordholm, K., Mitchell, W.: Memory effects in irreversible
  thermodynamics: Corrected derivation of transport equations.
\newblock Phys. Rev. A \textbf{5}(6), 2680 (1972)

\bibitem{sture1974strategies}
Sture, K., Nordholm, J., Zwanzig, R.: Strategies for fluctuation
  renormalization in nonlinear transport theory.
\newblock J. Stat. Phys. \textbf{11}(2), 143--158 (1974)

\bibitem{mori1973nonlinear}
Mori, H., Fujisaka, H.: On nonlinear dynamics of fluctuations.
\newblock Prog. Theor. Phys. \textbf{49}(3), 764--775 (1973)

\bibitem{kawasaki1973simple}
Kawasaki, K.: Simple derivations of generalized linear and nonlinear {L}angevin
  equations.
\newblock J. Phys. A \textbf{6}(9), 1289 (1973)

\bibitem{fujisaka1976fluctuation}
Fujisaka, H.: Fluctuation renormalization in nonlinear dynamics.
\newblock Prog. Theor. Phys. \textbf{55}(2), 430--437 (1976)

\bibitem{granek2021anomalous}
Granek, O., Kafri, Y., Tailleur, J.: The {A}nomalous {T}ransport of {T}racers
  in {A}ctive {B}aths.
\newblock arXiv preprint arXiv:2108.11970  (2021)

\bibitem{alder1970decay}
Alder, B.J., Wainwright, T.E.: Decay of the velocity autocorrelation function.
\newblock Phys. Rev. A \textbf{1}(1), 18 (1970)

\bibitem{pomeau1975time}
Pomeau, Y., Resibois, P.: Time dependent correlation functions and mode-mode
  coupling theories.
\newblock Phys. Rep. \textbf{19}(2), 63--139 (1975)

\bibitem{huang2011direct}
Huang, R., Chavez, I., Taute, K.M., Luki{\'c}, B., Jeney, S., Raizen, M.G.,
  Florin, E.L.: Direct observation of the full transition from ballistic to
  diffusive {B}rownian motion in a liquid.
\newblock Nat. Phys. \textbf{7}(7), 576--580 (2011)

\bibitem{franosch2011resonances}
Franosch, T., Grimm, M., Belushkin, M., Mor, F.M., Foffi, G., Forr{\'o}, L.,
  Jeney, S.: Resonances arising from hydrodynamic memory in {B}rownian motion.
\newblock Nature \textbf{478}(7367), 85--88 (2011)

\bibitem{kheifets2014observation}
Kheifets, S., Simha, A., Melin, K., Li, T., Raizen, M.G.: Observation of
  {B}rownian motion in liquids at short times: instantaneous velocity and
  memory loss.
\newblock Science \textbf{343}(6178), 1493--1496 (2014)

\bibitem{zwanzig1961memory}
Zwanzig, R.: Memory effects in irreversible thermodynamics.
\newblock Phys. Rev. \textbf{124}(4), 983 (1961)

\bibitem{saito2021microscopic}
Saito, K., Hongo, M., Dhar, A., Sasa, S.i.: Microscopic {T}heory of
  {F}luctuating {H}ydrodynamics in {N}onlinear {L}attices.
\newblock Phys. Rev. Lett. \textbf{127}(1), 010601 (2021)

\bibitem{d2016negative}
D'Alessio, L., Kafri, Y., Polkovnikov, A.: Negative mass corrections in a
  dissipative stochastic environment.
\newblock J. Stat. Mech. \textbf{2016}(2), 023105 (2016)

\bibitem{weinberg2017adiabatic}
Weinberg, P., Bukov, M., D’Alessio, L., Polkovnikov, A., Vajna, S.,
  Kolodrubetz, M.: Adiabatic perturbation theory and geometry of
  periodically-driven systems.
\newblock Phys. Rep. \textbf{688}, 1--35 (2017)

\bibitem{feng2021effective}
Feng, M., Hou, Z.: Effective dynamics of tracer in active bath: A mean-field
  theory study.
\newblock arXiv preprint arXiv:2110.00279  (2021)

\bibitem{nakayama2018unattainability}
Nakayama, Y., Kawaguchi, K., Nakagawa, N.: Unattainability of {C}arnot
  efficiency in thermal motors: {C}oarse graining and entropy production of
  {F}eynman-{S}moluchowski ratchets.
\newblock Phys. Rev. E \textbf{98}(2), 022102 (2018)

\bibitem{tailleur2008statistical}
Tailleur, J., Cates, M.E.: Statistical mechanics of interacting run-and-tumble
  bacteria.
\newblock Phys. Rev. Lett. \textbf{100}(21), 218103 (2008)

\bibitem{cover1999elements}
Cover, T.M., Thomas, J.A.: Elements of {I}nformation {T}heory, 2nd edn.
\newblock Wiley-Interscience, Hoboken, NJ (2006)

\bibitem{martin2021statistical}
Martin, D., O'Byrne, J., Cates, M.E., Fodor, {\'E}., Nardini, C., Tailleur, J.,
  van Wijland, F.: Statistical mechanics of active {O}rnstein-{U}hlenbeck
  particles.
\newblock Phys. Rev. E \textbf{103}(3), 032607 (2021)

\end{thebibliography}


\end{document}